\documentclass[twocolumn,pra,showkeys,showpacs,amsmath,amssymb,aps,superscriptaddress]{revtex4-1}

	\usepackage{bm}
	\usepackage{graphicx}
	\usepackage{amsmath}
	\usepackage{amssymb}
	\usepackage{mathrsfs}

	\usepackage[unicode=true,
 	bookmarks=false,
 	breaklinks=false,pdfborder={0 0 1},backref=false,colorlinks=true]
 	{hyperref}
	\hypersetup{
 	citecolor=blue}

	\usepackage{color}

	\usepackage{braket}
	
	\usepackage{float}

\graphicspath{ {./} }

\begin{document}

\title{Exact calculation of phonon effects on spin squeezing}

\author{D. Dylewsky}
\affiliation{Department of Physics, Georgetown University, 37$^{\rm th}$ and O Sts. NW, Washington, DC 20057 USA}

\author{J. K. Freericks}
\affiliation{Department of Physics, Georgetown University, 37$^{\rm th}$ and O Sts. NW, Washington, DC 20057 USA}

\author{M. L. Wall}
\affiliation{JILA, NIST, and University of Colorado, 440 UCB, Boulder, Colorado 80309, USA}

\author{A. M. Rey}
\affiliation{JILA, NIST, and University of Colorado, 440 UCB, Boulder, Colorado 80309, USA}

\author{M. Foss-Feig}
\affiliation{Joint Quantum Institute and the Joint Center for Quantum Information and Computer Science, NIST/University of Maryland, College Park, MD 20742 USA}

\begin{abstract}

Theoretical models of spins coupled to bosons provide a simple setting for studying a broad range of important phenomena in many-body physics, from virtually mediated interactions to decoherence and thermalization.  In many atomic, molecular, and optical systems, such models also underlie the most successful attempts to engineer strong, long-ranged interactions for the purpose of entanglement generation.  Especially when the coupling between the spins and bosons is strong---such that it cannot be treated perturbatively---the properties of such models are extremely challenging to calculate theoretically.  Here, exact analytical expressions for nonequilibrium spin-spin correlation functions are derived for a specific model of spins coupled to bosons.  The spatial structure of the coupling between spins and bosons is completely arbitrary, and thus the solution can be applied to systems in any number of dimensions.  The explicit and nonperturbative inclusion of the bosons enables the study of entanglement generation (in the form of spin squeezing) even when the bosons are driven strongly and near-resonantly, and thus provides a quantitative view of the breakdown of adiabatic elimination that inevitably occurs as one pushes towards the fastest entanglement generation possible.  The solution also helps elucidate the effect of finite temperature on spin squeezing. The model considered is relevant to a variety of atomic, molecular, and optical systems, such as atoms in cavities or trapped ions.  As an explicit example, the results are used to quantify phonon effects in trapped ion quantum simulators, which are expected to become increasingly important as these experiments push towards larger numbers of ions.

\end{abstract}


\maketitle

\section{Introduction \label{sec: intro}}

Spin-spin interactions play a crucial role in the generation of entanglement for applications in quantum information and metrology \cite{Horodecki:2009gb}.  In atomic, molecular, and optical (AMO) systems, intrinsic spin-spin couplings are often extremely weak, and generating entanglement much faster than decoherence timescales remains an important and challenging task.  One strategy to realize strong, long-ranged spin couplings, which is routinely employed in both trapped-ion systems and cavity QED, is to mediate them via a collection of auxiliary bosonic degrees of freedom ({\it e.\,g.\ }phonons in the case of trapped ions \cite{blatt-roos} and photons in cavity QED \cite{RevModPhys.73.565}).  If these bosonic modes are far off-resonance and the temperature is sufficiently low, they are only virtually occupied and can be (perturbatively) adiabatically eliminated \cite{PhysRevLett.92.207901}. This procedure yields approximate spin-only models that are generally easier to treat theoretically, and often more desirable experimentally.  For example, if $\omega$ is the characteristic energy input needed to create a boson, and $g$ is the characteristic coupling strength between the spins and bosons, spin-spin interactions of strength $\sim g^2/\omega$ can be generated.  However, the limit in which this procedure is quantitatively valid ($\omega\gg g$) is directly at odds with the limit in which the spin-dynamics is fastest (large $g^2/\omega$).  In order to overcome intrinsic timescale limitations, experiments are often forced to operate in parameter regimes where perturbative adiabatic elimination is not quantitatively justified, and a simple spin-only picture is questionable.

\begin{figure}[!t]
\includegraphics[width=0.75\columnwidth]{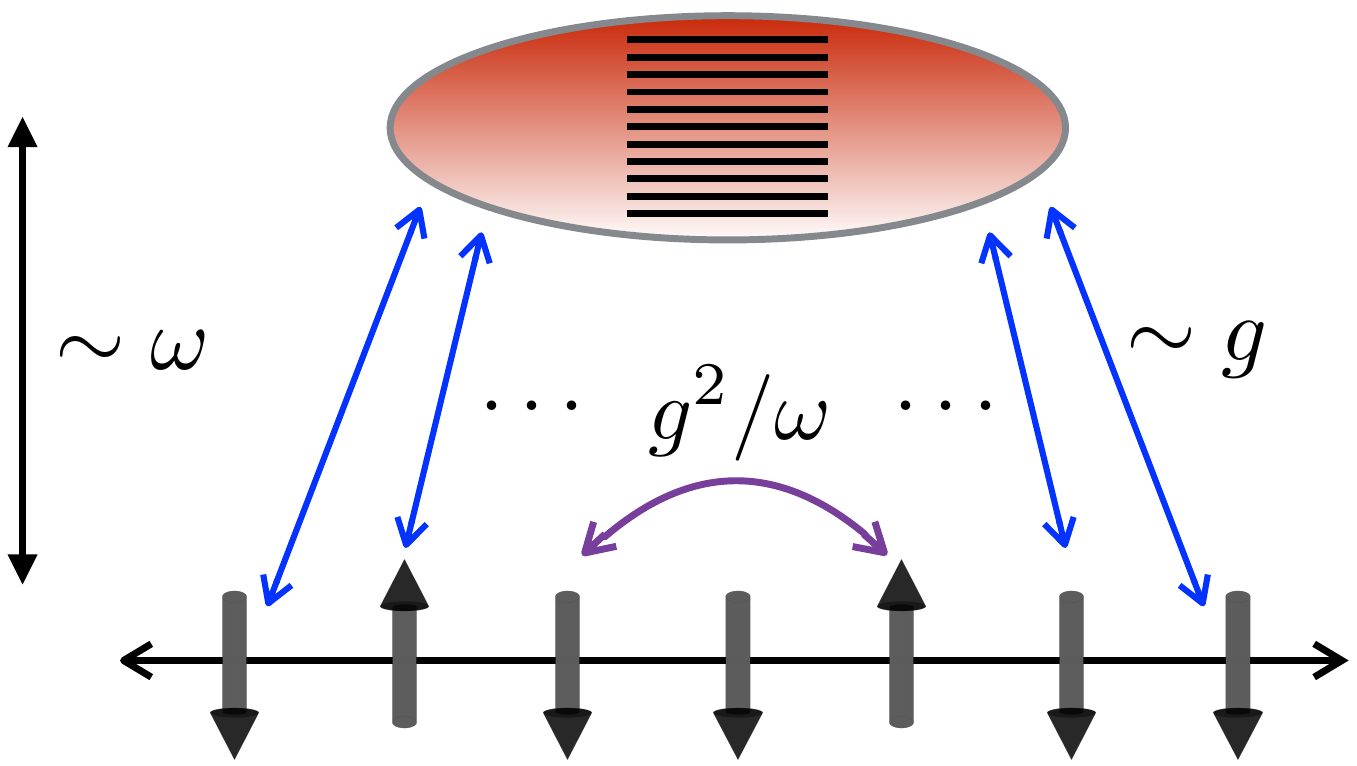}
\caption{(Color online) (a) Schematic of the model, with spins coupled (at characteristic coupling strength $g$) to a collection of noninteracting bosonic modes (at characteristic energy $\omega$).  When the phonons are far off-resonance, $\omega\gg g$, one can generally derive an approximate spin-only description of the system, which has direct spin-spin couplings of order $\sim g^2/\omega$.}
\label{fig:fig1}
\end{figure}

More generally, coupled spin-boson models play a central role in our understanding of quantum systems in contact with an environment, and have been studied extensively in both the condensed-matter and AMO communities for decades.  Even in the case of a single spin coupled to many noninteracting bosons \cite{RevModPhys.59.1}, or many spins coupled to a single bosonic mode \cite{PhysRev.93.99,gaudin}, remarkably rich and complex behavior emerges.  The general problem of many spins coupled to many bosons has very few analytically tractable limits, and is extremely difficult to study numerically, especially out-of-equilibrium and in more than one spatial dimension.  As such, exact solutions --- even of the simplest nontrivial models --- can play an important role in extending our understanding of these intricate coupled quantum systems.

Here, we provide an exact solution for the far-from-equilibrium dynamics of a collection of spins (with $S=1/2$) coupled uniaxially to a collection of noninteracting bosonic modes.  The solution is valid for arbitrary spatial structure of the bosonic modes, and therefore applies to systems in any number of spatial dimensions.  To a high degree of approximation, this model describes the dynamics of trapped-ion crystals when they are perturbed by a spin-dependent force \cite{P_Lee_2005}.  When the coupled spin-phonon system is driven far off (phonon) resonance, the phonons can be adiabatically eliminated and the dynamics is governed by an Ising Hamiltonian acting only on the spins.  Several experimental groups have exploited this result to engineer spin-entangled states of trapped ions \cite{sackett_2000}.  However, as system sizes increase, it is crucial to characterize and understand the discrepancies from this idealized situation that arise from finite population of the phonons, either due to their nonzero initial temperature or due to deviations from the far-off-resonance limit.  As a demonstration of its utility, the solution is used to calculate the spin squeezing generated dynamically by initializing the system in a product state of the spin and phonon degrees of freedom that is far from equilibrium.  The solution yields expressions that are efficient to evaluate numerically, enabling the calculation of dynamics for most experimentally achievable system sizes ($\mathcal{N}\lesssim 10^3$ spins and phonons).

The organization of the paper is as follows.  In Sec.\,\ref{sec:model} we present the model, and review its realization with trapped ions in a simple context. In Sec.\,\ref{sec:removing_time_ordering}, we explain the formalism used to derive the exact results for this system. Details for how
to explicitly calculate correlation functions are presented in Sec.\,\ref{sec:correlation_functions}, and numerical results for spin squeezing follow in Sec.\,\ref{sec: spin_squeezing}.  In Sec.\,\ref{sec:outlook} we discuss several interesting directions for future research.

\section{\label{sec:model}Model and its Realization with Ions}

The model we solve consists of a collection of $\mathcal{N}_{\rm b}$ bosonic modes coupled uniaxially to $\mathcal{N}_{\rm s}$ spins.  The spin-boson couplings can be time-dependent, and it is useful to break the Hamiltonian up into static and time varying parts as $\mathcal{H}(t)=\mathcal{H}_0+\mathcal{V}(t)$ \footnote{Here and throughout the rest of the manuscript, we work in units where $\hbar=1$.}, with
\begin{align}
\label{eq: Hamiltonian}
\mathcal{H}_0&=\sum_{\alpha=1}^{\mathcal{N}_{\rm b}}\omega_\alpha \hat{a}^\dagger_\alpha \hat{a}_\alpha^{\phantom\dagger},\\
\mathcal{V}(t)&=\sum_{j=1}^{\mathcal{N}_{\rm s}}\sum_{\alpha=1}^{\mathcal{N}_{\rm b}}\hat\sigma_j^z\left [ g_j^\alpha(t)\hat{a}^\dagger_\alpha + \bar g _j^\alpha(t)\hat{a}_\alpha\right ].\nonumber
\end{align}
Here, $\hat{a}^{\dagger}_{\alpha}$($\hat{a}^{\phantom\dagger}_{\alpha}$) creates (annihilates) a boson in a particular mode $\alpha$, and $\hat{\sigma}^r_j$ are the ($r=x,y,z$) Pauli spin matrices for the $j$$^{\rm th}$ spin.  The boson energies $\omega_{\alpha}$ in $\mathcal{H}_0$ are arbitrary, as are the coupling constants $g_j^{\alpha}(t)$ in $\mathcal{V}(t)$ (the overbar on $\bar{g}_j^{\alpha}(t)$ denotes complex conjugation).  A coupling to longitudinal fields $\sim\sum_{j}h_j\hat{\sigma}^z_j$ could also be included in $\mathcal{H}_0$, but such a term can be removed by working in a suitably rotating frame, and so it is ignored from the outset.  Also, note that terms coupling to spin directions other than $z$ are not included, and in general prohibit an exact solution.

For time-independent couplings (or, alternatively, at fixed $t$), the eigenstates of the above Hamiltonian are product states between all spins and suitably displaced vacuum states of the bosonic modes, and hence equilibrium properties of the model are essentially classical.  However, we are concerned with the response of the system when driven out of equilibrium; the ensuing relaxation dynamics is highly nontrivial, generically being accompanied by entanglement growth between the spins and bosons.  If the couplings $g_j^{\alpha}$ are independent of time, and if the interactions are weak ($g_{j}^{\alpha}\ll \omega_{\alpha})$, then $\mathcal{V}$ can be treated perturbatively.  For a system initialized in the boson vacuum, the bosons will only be populated virtually in the dynamics, which can therefore be described by an effective time-independent spin-only Hamiltonian.  For example, working to second order in $\mathcal{V}$, one obtains
\begin{align}
\label{eq: spin-spin hamiltonian}
\mathcal{H}_{\rm eff}^{\rm spin}=\sum_{j,k=1}^{\mathcal{N}_{\rm s}}J_{jk}\hat{\sigma}_j^z\hat{\sigma}_k^z~;~~~~~~J_{jk}=\sum_{\alpha=1}^{\mathcal{N}_{\rm b}}\frac{\bar{g}^{\alpha}_j g^{\alpha}_k}{\omega_{\alpha}}.
\end{align}
Even at this level of approximation, the spin dynamics is nontrivial, and exact time-dependent correlation functions were only recently obtained for general coupling constants $J_{jk}$ \cite{PhysRevA.87.042101,Kastner_NJP_2013,Foss-Feig_NJP_2013}. If the $J_{jk}$ do not depend on space, then Eq.\,(\ref{eq: spin-spin hamiltonian}) reduces to the single-axis-twisting model \cite{PhysRevA.47.5138}, which is a special case of the more general Lipkin-Meshkov-Glick model \cite{Lipkin_1965}.  In this case, the analysis is greatly simplified because the square of the total spin becomes a good quantum number, and the model can be solved in terms of collective spin variables \cite{PhysRev.93.99}.

\begin{figure}[t!]
\includegraphics[width=1.0\columnwidth]{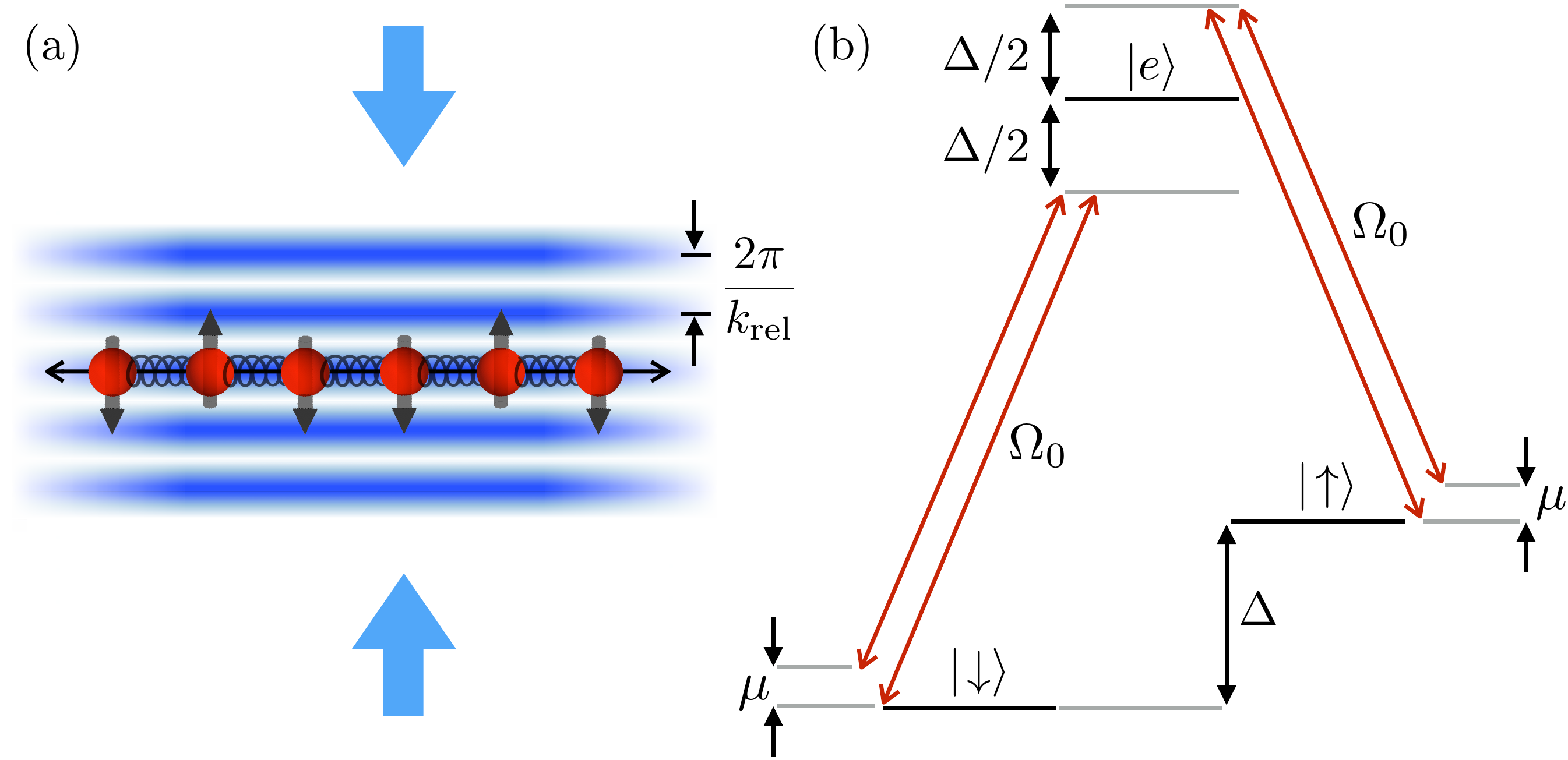}
\caption{(Color online) Realization of Eq.\,(\ref{eq: Hamiltonian}) with trapped ions: (a) Ions driven transversely via stimulated Raman transitions. The bosonic modes are realized as the normal modes of oscillation of the crystal around its equilibrium configuration (here shown as a 1D chain). (b) Simplified level diagram illustrating the essential ingredients for generating spin-phonon couplings in trapped ions.  Here, $\Omega_0$ denotes the strength of the coupling between the states $|\!\!\uparrow\rangle$,$|\!\!\downarrow\rangle$ and the optically-excited state $|e\rangle$ (figure not drawn to scale).}
\label{fig:level_diagram}
\end{figure}

Before solving for the time dependence of correlation functions induced by $\mathcal{H}(t)$, recall that Eq.\,(\ref{eq: Hamiltonian}) appears naturally in the description of various AMO systems.  For example, in ion traps the spin is realized by some internal structure of an ionized atom, and the bosons are excitations of the vibrational modes of the crystalized ions (phonons).  In cavity QED, where identical \cite{PhysRevA.75.013804} or closely related \cite{PhysRevA.81.021804} models can be realized, the spins are two-level neutral atoms and the bosons are photons in long-lived cavity modes.  In the context of trapped ions, the model can emerge in several different ways \cite{P_Lee_2005}, the conceptually simplest of which is through the application of a spin-dependent optical force to a crystal of ions \cite{leibfried2003,milburn2000,PhysRevA.62.022311} (though see Refs.\,\cite{PhysRevLett.82.1971,Islam03052013} for a common alternative realization). For example, in the spirit of Ref.\ \cite{britton12}, the ions can be driven by two lasers with difference frequency $\mu$ and relative wave-vector $\bm{k}_{\rm rel}$, as in Fig.\,\ref{fig:level_diagram}a.  Each ion is assumed to possess two long-lived hyperfine states labeled $|\!\!\uparrow\rangle$ and $|\!\!\downarrow\rangle$; they will represent the spin degree of freedom.  If the energy splitting between these two states is $\Delta$, then the ion Hamiltonian in the absence of driving is
\begin{equation}
\label{eq: no_drive}
\mathcal{H}^{\rm ion}=\sum_{\alpha=1}^{\mathcal{N}_{\rm b}}\omega_{\alpha}\hat{a}^{\dagger}_{\alpha}\hat{a}^{\phantom\dagger}_{\alpha}+\frac{\Delta}{2}\sum_{j=1}^{\mathcal{N}_{\rm s}}\hat{\sigma}^z_j.
\end{equation}
For simplicity, we assume that the crystal possesses a direction along which a single set of decoupled normal modes oscillate, the $z$-direction, and that $\bm{k}_{\rm rel}$ points along this direction; the index $\alpha$ in Eq.\,(\ref{eq: no_drive}) enumerates this set of modes. We also assume that the laser couples both spin states to a single optically-excited state $|e\rangle$, and choose the laser frequency so that the detunings from the optical transition are equal in magnitude and opposite in sign ($\pm\Delta/2$) for the two spin states (see Fig.\,\ref{fig:level_diagram}b). If the single-photon Rabi frequency $\Omega_0$ (assumed to be the same for both the $|\!\!\uparrow\rangle,|\!\!\downarrow\rangle\leftrightarrow|e\rangle$ transitions) is small compared to the single-photon detuning $\Delta/2$, the electronic excited state can be adiabatically eliminated, leaving behind an AC-Stark shift for each spin state that oscillates in time at the difference frequency $\mu$ and in space at the difference wave vector ${\bm k}_{\rm rel}$.  Combined with a rotating-wave approximation ({\it i.\,e.}~ignoring all terms with optical-frequency time-dependences) and a frame transformation to remove the energy splitting $\Delta$, adiabatic elimination of $|e\rangle$ yields 
\begin{equation}
\label{eq: ion_non_lamb_dicke}
\mathcal{H}_{\rm driven}^{\rm ion}(t)=\sum_{\alpha=1}^{\mathcal{N}_{\rm b}}\omega_{\alpha}\hat{a}^{\dagger}_{\alpha}\hat{a}^{\phantom\dagger}_{\alpha}+\Omega\sum_{j=1}^{\mathcal{N}_{\rm s}}\cos(k_{\rm rel}\hat{z}_j-\mu t)\hat{\sigma}_{j}^z.
\end{equation}
Here, $\Omega\equiv 4\Omega_0^2/\Delta$ is the characteristic strength of the spin-dependent AC-Stark shift experienced by the states $|\!\!\!\uparrow\rangle$ and $|\!\!\!\downarrow\rangle$, and $k_{\rm rel}=|\bm{k}_{\rm rel}|$.  The position operator along the $z$-direction for the $j$$^{\rm th}$ ion, denoted $\hat z_j$, can be expanded in terms of creation and annihilation operators for the normal modes of the crystal as $k_{\rm rel}\hat{z}_j=\sum_{\alpha}\eta_{\alpha}b_{j}^{\alpha}(\hat{a}^{\dagger}_{\alpha}+\hat{a}^{\phantom\dagger}_{\alpha})$.  Here, $b_{j}^{\alpha}$ is the orthogonal normal-mode transformation matrix, and $\eta_{\alpha}=k_{\rm rel}\sqrt{\hbar/(2m\omega_{\alpha})}$ (restoring $\hbar$ temporarily) parametrizes how small the characteristic ion displacements in the ground state of the mode $\alpha$ are compared to the length scale $k_{\rm rel}^{-1}$ over which the applied spin-dependent potential changes appreciably.  In the Lamb-Dicke limit, $\eta_{\alpha}\ll 1$ for all $\alpha$, and working to lowest order in $\eta_{\alpha}$, Eq.\,(\ref{eq: ion_non_lamb_dicke}) becomes
\begin{align}
\label{eq: hamiltonian_for_ions}
\mathcal{H}^{\rm ion}_{\rm driven}(t)&\approx \sum_{\alpha=1}^{\mathcal{N}_{\rm b}}\omega_{\alpha}
\hat{a}^{\dagger}_{\alpha}\hat{a}^{\phantom\dagger}_{\alpha} \\
&+\Omega\sin(\mu t)\sum_{j=1}^{\mathcal{N}_{\rm s}}\sum_{\alpha=1}^{\mathcal{N}_{\rm b}}\hat{\sigma}^z_j\eta_{\alpha}b_{j}^{\alpha}(a^{\dagger}_{\alpha}+a^{\phantom\dagger}_{\alpha}),\nonumber
\end{align}
which is Eq.\,(\ref{eq: Hamiltonian}) with $g_{j}^{\alpha}(t)=\bar{g}_{j}^{\alpha}(t)=\Omega\sin(\mu t)\eta_{\alpha}b^{\alpha}_j$.  Having motivated the general form of the Hamiltonian in Eq.\,(\ref{eq: Hamiltonian}), we now proceed to compute correlation functions evolving under it.  With the formal solution in hand, however, we will eventually return to the context of trapped ions and Eq.\,(\ref{eq: hamiltonian_for_ions}) when discussing the application of our results to computing spin squeezing in Sec.\ \ref{sec: spin_squeezing}.

\section{Solution for Correlation Functions}



The following section includes technical derivations that are not essential for following most of the discussion in Sec.\,\ref{sec: spin_squeezing}; readers wishing to skip these details can proceed directly to that section.  Because $\mathcal{V}(t)$ in Eq.\,(\ref{eq: Hamiltonian}) is explicitly time-dependent, the time-evolution operator corresponding to $\mathcal{H}(t)$ must be written as a time-ordered product, which complicates the calculation of observables.  The first step in obtaining closed forms for correlation functions, therefore, is to obtain an explicit form of the time evolution operator that does \emph{not} require time ordering.  It is well known (c.f.\ Refs.\ \cite{PhysRevLett.91.187902,J_Wang_2012}) that this can be accomplished via appropriate factorizations of the time-evolution operator.  However, in the interest of maintaining a self-contained solution of the model, this procedure is briefly reviewed in Sec.\,\ref{sec:removing_time_ordering}.  With an explicit form of the time-evolution operator in hand, we then move on to our main formal results in Sec.\,\ref{sec:correlation_functions}, obtaining closed-form expressions for spin-spin correlation functions.

\subsection{\label{sec:removing_time_ordering}Explicit form for the time-evolution operator}

The time-evolution operator satisfies the equation of motion $i\partial\mathcal{U}(t,t_0)/\partial t=\mathcal{H}(t)\mathcal{U}(t,t_0)$ with respect to the full Hamiltonian $\mathcal{H}(t)$ defined in Eq.\,(\ref{eq: Hamiltonian}), and can be written as a time-ordered product,
\begin{equation}
\mathcal{U}(t,t_0)=\mathcal{T}_t \exp \bigg(-i\int_{t_0}^t d\tau \mathcal{H}( \tau)\bigg),
\label{eq: udef}
\end{equation}
with $\mathcal{T}_t$ the time-ordering operator. The first step in rewriting the time-evolution operator without the need for time ordering is to move to the interaction picture with
respect to $\mathcal{H}_0$.  Defining the ``perturbation'' in the interaction picture,
\begin{align}
&\mathcal{V}_I(t,t_0)=e^{i\mathcal{H}_0(t-t_0)}\mathcal{V}(t)e^{-i\mathcal{H}_0(t-t_0)}\label{eq: v_int}\\
&=\sum_{j=1}^{\mathcal{N}_{\rm s}}\sum_{\alpha=1}^{\mathcal{N}_{\rm b}}
\hat\sigma_j^z\left [ g_j^\alpha(t)e^{i\omega_{\alpha}(t-t_0)}\hat{a}^\dagger_\alpha + \bar g _j^\alpha(t)e^{-i\omega_{\alpha}(t-t_0)}\hat{a}_\alpha\right ],\nonumber
\end{align}
we can write $\mathcal{U}(t,t_0)=e^{-i\mathcal{H}_0(t-t_0)}\mathcal{U}_{I}(t,t_0)$, with
\begin{equation}
\mathcal{U}_I(t,t_0)\equiv \mathcal{T}_t
\exp\bigg(-i\int_{t_0}^t d\tau \mathcal{V}_I(\tau,t_0)\bigg).
\label{eq: int_rep_fact}
\end{equation}
Here $\mathcal{H}_0(t-t_0)$ is the product of the ``unperturbed'' Hamiltonian $\mathcal{H}_0$ and the time difference $t-t_0$; the latter should not be confused as an argument of $\mathcal{H}_0$, which is manifestly time-independent.

The next step follows the textbook problem of driven harmonic oscillators \cite{Landau_Lifshitz,Gottfried}.  Defining the operator
\begin{equation}
\mathcal{W}(t,t_0) \equiv \int_{t_0}^t d\tau \mathcal{V}_I(\tau, t_0),
\label{eq: wdef}
\end{equation}
which satisfies $d\mathcal{W}(t,t_0)/dt=\mathcal{V}_I(t,t_0)$, we further factorize $\mathcal{U}_I(t,t_0)=e^{-i \mathcal{W}(t,t_0)}\tilde{\mathcal{U}}(t,t_0)$, with
\begin{align}
\tilde{\mathcal{U}}(t,t_0)\equiv e^{i \mathcal{W}(t,t_0)}\mathcal{U}_I(t,t_0).
\end{align}
The benefit of this factorization becomes immediately clear upon differentiating $\tilde{\mathcal{U}}(t,t_0)$ with respect to $t$. We take the derivative with the help of the equality
\begin{align}
\label{eq:simple_identity}
\frac{d e^{i\mathcal{W}(t,t_0)}}{dt}=\left.\left(\frac{d}{d\lambda}e^{i[\mathcal{W}(t,t_0)+\lambda \mathcal{V}_{I}(t,t_0)]}\right)\right|_{\lambda=0},
\end{align}
which can be verified by Taylor-expanding both sides.  Crucially, the commutator of $\mathcal{W}(t_2,t_0)$ with $\mathcal{V}_I(t_1,t_0)$ depends only on the operators $\hat{\sigma}_j^z$, and hence commutes with both $\mathcal{W}$ and $\mathcal{V}_I$ at all other times,
\begin{align}
\label{eq:triple_identity_1}
\left [\mathcal{W}(t_3,t_0),\left [ \mathcal{W}(t_2,t_0),\mathcal{V}_I(t_1,t_0)\right ] \right ]&=0\\
\label{eq:triple_identity_2}
\left [\mathcal{V}_{I}(t_3,t_0),\left [ \mathcal{W}(t_2,t_0),\mathcal{V}_I(t_1,t_0)\right ] \right ]&=0.
\end{align}
As an immediate consequence of Eqs.\,(\ref{eq:triple_identity_1}) and (\ref{eq:triple_identity_2}), one can make the replacement
\begin{align}
e^{i[\mathcal{W}(t,t_0)+\lambda \mathcal{V}_I(t,t_0)]}=e^{i\mathcal{W}(t,t_0)}e^{i\lambda \mathcal{V}_I(t,t_0)}e^{\lambda[\mathcal{W}(t,t_0),\mathcal{V}_I(t,t_0)]/2}\nonumber
\end{align}
in Eq.\,(\ref{eq:simple_identity}), evaluate the right-hand-side, and thereby obtain
\begin{align}
\label{eq:U_tilde_difeq}
\frac{d}{dt}\tilde{\mathcal{U}}(t,t_0)=\frac{1}{2}[\mathcal{W}(t,t_0),\mathcal{V}_I(t,t_0)]\tilde{\mathcal{U}}(t,t_0).
\end{align}
Because Eqs.\,(\ref{eq:triple_identity_1}) and (\ref{eq:triple_identity_2}) hold for all times, Eq.\,(\ref{eq:U_tilde_difeq}) can be integrated \emph{without} regard for time ordering, yielding $\tilde{\mathcal{U}}(t,t_0)=\exp(\int_{t_0}^{t}d\tau [\mathcal{W}(\tau,t_0),\mathcal{V}_I(\tau,t_0)]/2)$.  The full time-evolution operator can now be written as
\begin{align}
\mathcal{U}(t,t_0)&=e^{-i\mathcal{H}_0 (t-t_0)} e^{-i\mathcal{W}(t,t_0)}\nonumber\\
&\times\exp \bigg( \frac{1}{2} \int_{t_0}^t d\tau \left [\mathcal{W}(\tau,t_0),\mathcal{V}_I(\tau,t_0)\right ]\bigg).
\label{eq: u_final}
\end{align}
At this point we have reduced the evaluation of a time-ordered product to the evaluation of the product of three different time-evolution terms, each generated by an operator that commutes with itself at different times [but note that in general the order of the three exponential factors in Eq.\,(\ref{eq: u_final}) must be maintained]. It turns out that for the choice of Hamiltonian in Eq.\,(\ref{eq: Hamiltonian}), only the first and second factor do not commute with each other, and need to maintain their relative ordering.

The operator $\mathcal{W}(t,t_0)$ can be written explicitly as
\begin{equation}
\mathcal{W}(t,t_0) = i\sum_{j=1}^{\mathcal{N}_{\rm s}}\sum_{\alpha=1}^{\mathcal{N}_{\rm b}}
\left [ A_j^\alpha(t,t_0)\hat a_\alpha^\dagger-\bar{A}^\alpha_j(t,t_0)\hat a_\alpha^{\phantom\dagger}\right ]
\hat\sigma_j^z,
\end{equation}
with
\begin{align}
A_j^\alpha(t,t_0) = -i\int_{t_0}^t d\tau g_j^\alpha(\tau)e^{i\omega_\alpha (\tau-t_0)}.
\label{eq: A_def_general}
\end{align}
Taking the commutator in the third factor of Eq.\,(\ref{eq: u_final}) and integrating yields
\begin{align}
\label{eq: U_final}
\mathcal{U}(t,t_0)&=\exp \left [ -i\mathcal{H}_0 (t-t_0)\right] \exp [-i\mathcal{W}(t,t_0)]\nonumber\\
&\times\exp\big[ -i\sum_{j,k=1}^{\mathcal{N}_{\rm s}} \mathcal{S}_{jk}(t,t_0)\hat \sigma_j^z\hat\sigma_{k}^z\big],
\end{align}
with
\begin{align}
\label{eq: S_def_general}
&\mathcal{S}_{jk}(t,t_0)=\\
&{\rm Im}\!\sum_{\alpha=1}^{\mathcal{N}_{\rm b}}\int_{t_0}^t \!\!\!d\tau
\!\!\int_{t_0}^{\tau} \!\!\!\!d\tau' \! \left(\!\frac{g_j^\alpha(\tau')\bar{g}_{k}^\alpha(\tau)+g_k^\alpha(\tau')\bar{g}_{j}^\alpha(\tau)}{2}\!\right)\! e^{i\omega_\alpha(\tau'-\tau)}.\nonumber
\end{align}
Note that we have utilized $[\hat{\sigma}^z_j,\hat{\sigma}^z_k]=0$ to write the Ising coefficients in an explicitly symmetric form, so $\mathcal{S}_{jk}=\mathcal{S}_{kj}$.

\subsection{\label{sec:correlation_functions}Calculating time-dependent expectation values of spin operators}

To simplify the notation in what follows, we set $t_0=0$ and suppress its appearance, in which case the time-dependent expectation value of an operator $\hat{\mathcal{O}}$ is given by $\mathcal{O}(t)=\bra{\psi_0}\mathcal{U}^\dagger(t)\hat{\mathcal{O}}\,\mathcal{U}(t)\!\ket{\psi_0}$, with $\ket{\psi_0}$ the initial wave function at $t=0$ and the evolution operator given in Eq.\,(\ref{eq: U_final}).  We will only consider expectation values of spin operators; because they commute 
with the bosonic Hamiltonian $\mathcal{H}_0$, the boson-only part of the evolution operator always cancels out, leaving
\begin{align}
\label{eq: expectation_value}
\mathcal{O}(t)&=\bra{\psi(t)}\hat{\mathcal{O}}\ket{\psi(t)},\\
\ket{\psi(t)}&=e^{-i\mathcal{W}(t)}e^{-i\sum_{j,k=1}^{\mathcal{N}_{\rm s}} \mathcal{S}_{jk}(t)\hat \sigma_j^z\hat\sigma_{k}^z}\ket{\psi_0}.\nonumber
\end{align}

The results that follow can easily be worked out for arbitrary product states between all of the degrees of freedom (spin and boson).  However, in order to simplify the discussion we present results only for initial states where all spins initially point along the $+x$ axis,
\begin{align}
\label{eq: initial_state}
\ket{\psi_0}=2^{-\mathcal{N}_{\rm s}/2}\!\!\!\!\!\!\sum_{\sigma_1,...,\sigma_{\mathcal{N}_{\rm s}=\uparrow,\downarrow}}\!\!\!\!\!\ket{\sigma_1}\otimes\cdots\otimes\ket{\sigma_{\mathcal{N}_{\rm s}}}\bigotimes_{\alpha=1}^{\mathcal{N}_{\rm b}} \ket{\varphi_\alpha}.
\end{align}
The bosons are taken to be in a product state between the different modes, but we allow the state of any particular mode, $\ket{\varphi_{\alpha}}$, to be arbitrary. While Eq.\,(\ref{eq: initial_state}) may seem restrictive, it is a natural choice for the generation of spin squeezing, as will be made clear in Sec.\,\ref{sec: spin_squeezing}.

To characterize the spin dynamics of the system, we compute a number of time-dependent expectation values of products of spin operators.  Defining spin raising (+) and lowering (-) operators $\hat \sigma_j^\pm=(\hat \sigma_j^x\pm i\hat\sigma_j^y)/2$ for each spin $j$, we compute
\begin{align}
\label{eq: cf_1}
\langle\hat\sigma^a_m\rangle&=\bra{\psi(t)}\hat\sigma^a_m\ket{\psi(t)},\\
\label{eq: cf_2}
\langle\hat\sigma^a_m\hat\sigma^z_n\rangle&=\bra{\psi(t)}\hat\sigma^a_m\hat\sigma^z_n\ket{\psi(t)},\\
\label{eq: cf_3}
\langle\hat\sigma^a_m\hat\sigma^b_n\rangle&=\bra{\psi(t)}\hat\sigma^a_m\hat\sigma^b_n\ket{\psi(t)},
\end{align}
with $a,b=\pm$ and $m\neq n$.  Note that all nontrivial expectation values of one or two spin operators can be obtained as linear combinations of those in Eqs.\,(\ref{eq: cf_1}-\ref{eq: cf_3}) (correlation functions involving only the $\hat{\sigma}^z_j$ are independent of time, as $[\mathcal{H}(t),\hat{\sigma}_j^z]=0$).

In order to calculate these expectation values, we make a few observations about Eq.\,(\ref{eq: expectation_value}). First, the final factor of the time-evolution operator, which involves only the Ising spin operators, can be further factorized into a product of exponentials (instead of an exponential of a sum of terms), because each term in the exponent commutes with every other term in the exponent. Second, we can factorize the exponential of $\mathcal{W}(t)$ into factors that have fixed $\alpha$ (but still sum over $j$), because those operators also commute with each other. Each one of these factors that commutes with the operator $\hat{\mathcal{O}}$ can be moved from the right, through $\hat{\mathcal{O}}$, and then cancels against an inverse term on the left coming from hermitian conjugate of the time-evolution operator.

Using the identity
\begin{equation}
e^{i\lambda\hat\sigma^z_m}\hat\sigma^\pm_m e^{-i\lambda\hat\sigma^z_m}
= \hat\sigma_m^\pm e^{\pm 2i\lambda},
\end{equation}
valid for $[\lambda,\hat{\sigma}_m^z]=[\lambda,\hat{\sigma}_m^{\pm}]=0$, together with
\begin{align}
&e^{i\mathcal{W}(t,t_0)}\hat\sigma^\pm_m e^{-i\mathcal{W}(t,t_0)}\\
&=e^{i\mathcal{W}(t,t_0)} e^{\mp 2 \sum_{\alpha=1}^{\mathcal{N}_{\rm b}}\left [
A_m^\alpha(t)\hat a^\dagger_\alpha - \bar{A}_m^{\alpha}(t)\hat a^{\phantom\dagger}_\alpha\right ]}
e^{-i\mathcal{W}(t,t_0)}\hat\sigma_m^\pm\nonumber\\
&=e^{\mp2\big(\sum_{\alpha=1}^{\mathcal{N}_{\rm b}}\left [
A_m^\alpha(t)\hat a^\dagger_\alpha - \bar{A}_m^{\alpha}(t)\hat a^{\phantom\dagger}_\alpha\right ]-\sum_{j\neq m}^{\mathcal{N}_{\rm s}}{\rm Im}\left [ \bar{A}_m^{\beta}(t)A_j^{\beta}(t)\right] \hat\sigma_j^z\big)}\hat\sigma_m^\pm,\nonumber
\end{align}
which follow from standard operator identities, we can simplify the expectation values of the operator averages that we are calculating.  After some additional algebra, and defining the displacement operators $\hat{D}_{\alpha}(\vartheta)=\exp(\vartheta \hat{a}_{\alpha}^{\dagger}-\bar{\vartheta} \hat{a}^{\phantom\dagger}_{\alpha})$ and modified spin-spin couplings
\begin{align}
\tilde{\mathcal{S}}_{mn}(t)=\mathcal{S}_{mn}(t)+\frac{1}{2}{\rm Im}\sum_{\beta=1}^{\mathcal{N}_{\rm b}}\bar{A}_m^{\beta}(t)A_n^{\beta}(t),
\end{align}
the final results are as follows:
\begin{align}
\label{eq: a_expectation}
\langle\hat{\sigma}_m^a\rangle = \frac{1}{2}\prod_{\alpha =1}^{\mathcal{N}_{\rm b}}
\!\bra{\varphi_\alpha}\hat{D}^{\dagger}_{\alpha}\big(2aA_m^\alpha(t)\big)\ket{\varphi_\alpha}\! \prod_{j\ne m}^{\mathcal{N}_{\rm s}} \!\cos \big(4\,\tilde{\mathcal{S}}_{mj}(t)\big)
\end{align}
\begin{align}
\label{eq: az_correlations}
\left\langle\hat{\sigma}_m^a\hat{\sigma}_n^z\right\rangle &= \frac{ai}{2}\prod_{\alpha=1}^{\mathcal{N}_{\rm b}} \bra{\varphi_\alpha}\hat{D}^{\dagger}_{\alpha}\big(2aA_m^\alpha(t)\big)\ket{\varphi_\alpha}\\
&\times\sin \big(4\,\tilde{\mathcal{S}}_{mn}(t)\big)\prod_{j\neq m, n}^{\mathcal{N}_{\rm s}}\cos \big(4\,\tilde{\mathcal{S}}_{mj}(t)\big).\nonumber
\end{align}
\begin{align}
\label{eq: ab_correlations}
\left\langle\hat{\sigma}_m^a\hat{\sigma}_n^b\right\rangle &=\frac{1}{4}\prod_{\alpha=1}^{\mathcal{N}_{\rm b}}\bra{\varphi_\alpha} \hat{D}^{\dagger}_{\alpha}\big(2a A_m^\alpha (t) + 2b A_n^\alpha (t)\big)\ket{\varphi_\alpha}\nonumber\\
&\times\prod_{j\neq m, n}^{\mathcal{N}_{\rm s}}\cos \big(4a\tilde{\mathcal{S}}_{mj}(t)+4b\tilde{\mathcal{S}}_{nj}(t)\big)
\end{align}
From Eq.\,(\ref{eq: A_def_general}), we see that ${\rm Im} [\bar{A}_m^{\alpha}(t)A_n^\alpha(t)]=0$, and therefore $\tilde{\mathcal{S}}_{mn}(t)=\mathcal{S}_{mn}(t)$, whenever $\bar{g}_m^{\alpha}(t_1)g_{n}^{\alpha}(t_2)=\bar{g}_n^{\alpha}(t_1)g_{m}^{\alpha}(t_2)$.  This situation is realized  for the normal modes of ions in linear Paul traps, and also for the axial modes of Penning traps (though not the in-plane modes) \cite{PhysRevA.87.013422}.  There, $\bar{g}_j^{\alpha}(t)=g_{j}^{\alpha}(t)=\Omega\sin(\mu t)\eta_{\alpha}b^{\alpha}_j$, since the normal-mode transformation matrix $b^{\alpha}_j$ can always be chosen to be real.

\subsection{Evaluation of the boson matrix elements\label{sec:phonon_matrix_elements}}
A boson matrix element of the form $\bra{\varphi_\alpha}\hat{D}_{\alpha}(\vartheta)\ket{\varphi_\alpha}$ can easily be evaluated for an arbitrary state $\ket{\varphi_\alpha}=\sum_{n=0}^\infty c_n^\alpha\ket{n}_\alpha$, where $\ket{n}_\alpha = \frac{1}{\sqrt{n!}}\left ( \hat{a}^{\dagger}_\alpha\right )^n\ket{0}_\alpha$ are normalized Fock states of the $\alpha$$^{\rm th}$ mode.  Writing the displacement operator as $\hat{D}_{\alpha}(\vartheta)=e^{-|\vartheta|^2/2}e^{\vartheta \hat{a}^{\dagger}_{\alpha}}e^{-\bar{\vartheta}\hat{a}^{\phantom\dagger}_{\alpha}}$, straightforward algebra leads to
\begin{align}
\label{eq: displacement_ev}
\bra{\varphi_\alpha}\hat{D}_{\alpha}(\vartheta)\ket{\varphi_\alpha}=\!\!\!\sum_{n,p,q=0}^\infty\!\!\!\frac{\bar{c}^{\alpha}_{n+p}c^\alpha_{n+q}\vartheta^p\bar{\vartheta}^q\sqrt{(n+q)!(n+p)!}}{p!q!n! e^{|\vartheta|^2/2}}.
\end{align}
The complete boson matrix elements in Eqs.\,(\ref{eq: a_expectation}-\ref{eq: ab_correlations}) follow by taking the product of Eq.\,(\ref{eq: displacement_ev}) over all of the $\mathcal{N}_{\rm b}$ modes labeled by $\alpha$. Expectation values such as Eq.\,(\ref{eq: displacement_ev}) can easily be generalized to deal with finite temperature states of the bosonic modes.  For example, in what follows we consider a situation where all bosonic modes are thermally populated at an inverse temperature $\beta$, such that the initial boson density matrix is $\varrho(\beta)=\bigotimes_\alpha\rho_{\alpha}(\beta)$, with
\begin{align}
\label{eq: density_matrix}
\rho_{\alpha}(\beta)&=\mathcal{Z}_{\alpha}(\beta)^{-1}\sum_{n_{\alpha}=0}^{\infty}\ket{n_{\alpha}}\bra{n_{\alpha}}e^{-\beta \omega_{\alpha}n_{\alpha}},\\
\mathcal{Z}_{\alpha}(\beta)&=\sum_{n_{\alpha}=0}^{\infty}e^{-\beta \omega_{\alpha}n_{\alpha}}=\frac{1}{1-e^{-\beta\omega_{\alpha}}}.
\end{align}
Expectation values of the form $\bra{\varphi_{\alpha}}\hat{D}_{\alpha}(\vartheta_{\alpha})\ket{\varphi_{\alpha}}$ appearing in Eqs.\,(\ref{eq: a_expectation}-\ref{eq: ab_correlations}) should be replaced with $\mathscr{D}_{\alpha}(\beta,\vartheta_{\alpha})\equiv{\rm Tr}\big[\rho_{\alpha}(\beta)\hat{D}_{\alpha}(\vartheta_{\alpha})\big]$. Inserting Eq.\,(\ref{eq: density_matrix}) into this trace and utilizing Eq.\,(\ref{eq: displacement_ev}), straightforward algebra leads to
\begin{align}
\label{eq: finite_T_matrix_element}
\mathscr{D}_{\alpha}(\beta,\vartheta_{\alpha})=\exp\left(-\frac{1}{2}|\vartheta_{\alpha}|^2\coth(\beta\omega_{\alpha}/2)\right).
\end{align}

\section{ \label{sec: spin_squeezing}Spin Squeezing}

It is also possible to obtain closed-form expressions for higher-order ($n$-spin) versions of the $2$-spin correlation functions derived above.  However, already at the level of $2$-spin correlation functions, we can learn a great deal about the time evolution of the system, and the nature of the entanglement that develops.  For example, from the correlation functions in Eqs.\,(\ref{eq: az_correlations}) and (\ref{eq: ab_correlations}), we can compute the variance of the spin distribution, which enables us to characterize spin squeezing \cite{PhysRevA.47.5138,ma_2011}.  Spin squeezing is just one of many measures of entanglement in a many-body system, which has the virtue of quantifying the potential enhancement in precision obtainable in Ramsey spectroscopy (as compared to the case of unentangled spins) \cite{PhysRevA.46.R6797}.  Moreover, it establishes a lower bound on the depth of entanglement, {\it i.\,e.}\ the minimum number of simultaneously entangled particles in the system \cite{PhysRevLett.86.4431}.

\begin{figure}
	\centering
	\includegraphics[width=0.9\columnwidth]{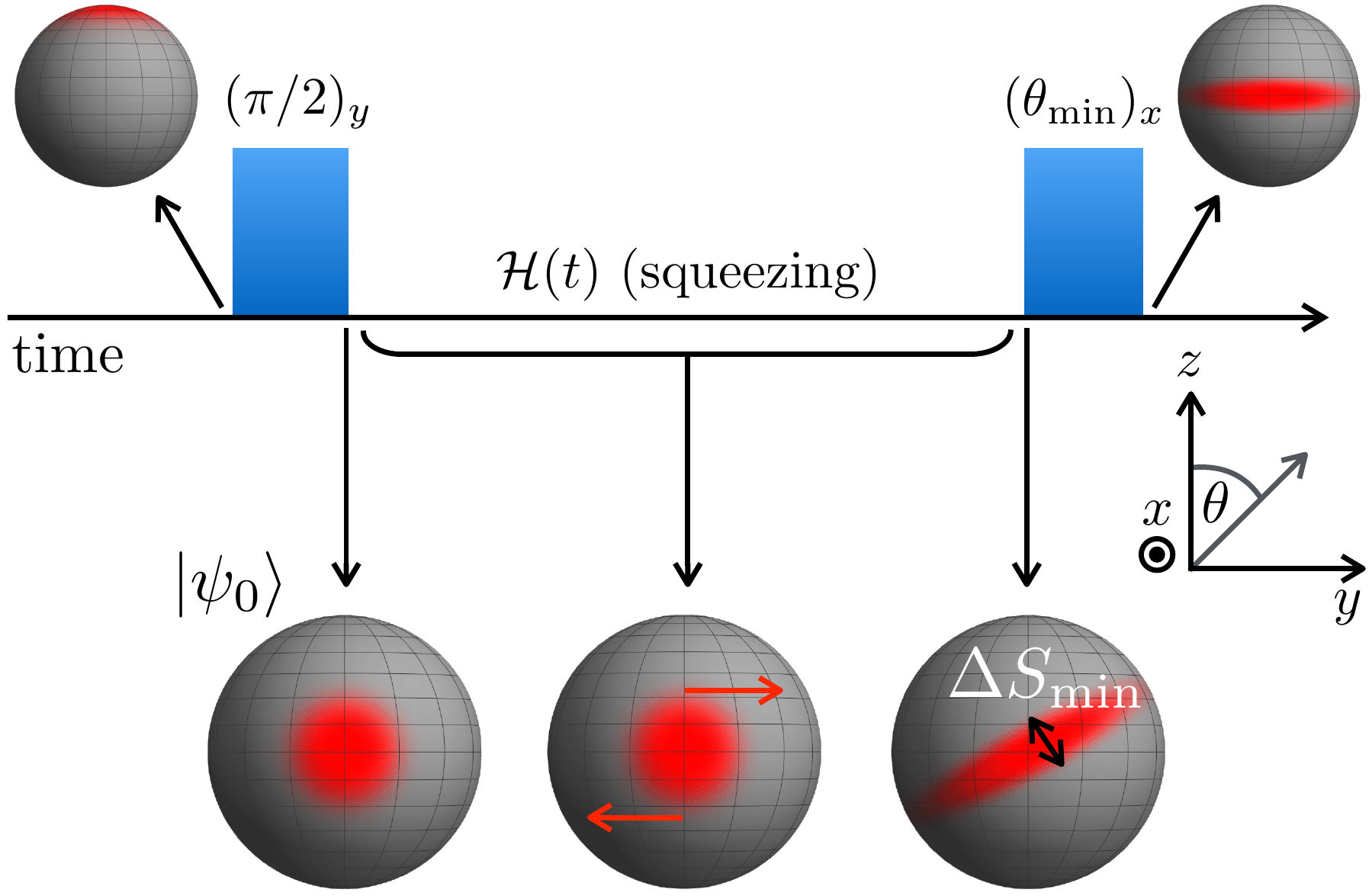}
	\label{bloch_sphere_diagrams}
	\caption{(Color online) Illustration of an experimental protocol employed to generate spin squeezing.  Spins are forced into a nonequilibrium state polarized along the $+x$ axis.  The Hamiltonian then acts for some period of time, causing squeezing of the initially Gaussian spin state, after which the minimal variance of the spin distribution is mapped onto the $z$-axis and measured.  The coordinate system shown corresponds to that used in Eq.\,(\ref{eq: squeeze_def}).}
\end{figure}

If the bosonic modes are initially cooled to a vacuum state, driven weakly so that they can be adiabatically eliminated, and if the resulting effective spin-spin coupling strength $J$ is independent of the spatial distance between the spins, the dynamics is governed by the single-axis-twisting Hamiltonian $\mathcal{H}_{\rm sat}=4J\hat{S}_z\hat{S}_z$, where $\hat{S}_z\equiv(1/2)\sum_{j}\hat{\sigma}_j^z$ \cite{PhysRevA.47.5138}.  In this model, spin squeezing is generated by first polarizing the collective spin vector along the $+x$ axis, and then letting it evolve under $\mathcal{H}_{\rm sat}$.  Making a mean-field approximation, $\mathcal{H}_{\rm sat}\approx 8J\langle\hat{S}_z\rangle\hat{S}_z-4J\langle\hat{S}_z\rangle^2$, the dynamics can be understood as a precession about the $z$-axis in a direction determined by the mean $z$-component of the spin; the initial spin state has quantum fluctuations above and below the equator, and therefore this dynamics causes it to get sheared and elongated, as in Fig.\,\ref{bloch_sphere_diagrams}.   Uncertainty along one axis is reduced (squeezed), while uncertainty in an orthogonal direction is increased.

The extent of squeezing along a particular direction in the $yz$ plane can be quantified by the parameter
\begin{equation}
\label{eq: squeeze_def}
\xi(\theta) = \mathcal{N}_{\rm s}^{\frac{1}{2}}\frac{\Delta S_\theta}{|\langle\hat{S}_x\rangle|},
\end{equation}
where $\hat{S}_{\theta}\equiv\hat{S}_z\cos\theta\!+\!\hat{S}_y\sin\theta$ and $\Delta S_\theta=(\langle\hat{S}_{\theta}^2\rangle\!-\!\langle\hat{S}_{\theta}\rangle^2)^{\frac{1}{2}}$.  The spin-squeezing parameter is then defined by minimizing the standard deviation, $\Delta S_{\rm min}=\min_{\theta}\Delta S_{\theta}$, such that $\xi=\min_{\theta}\xi(\theta)=\mathcal{N}_{\rm s}^{1/2}\Delta S_{\rm min}/|\langle\hat{S}_x\rangle|$.  Straightforward algebra enables the optimal angle to be expressed explicitly in terms of spin-spin correlation functions \cite{ma_2011},
\begin{equation}
\theta = \frac{1}{2}\arctan\left(\frac{\langle \hat{S}_z \hat{S}_y \rangle + \langle \hat{S}_y \hat{S}_z \rangle} {\langle \hat{S}_z \hat{S}_z \rangle - \langle \hat{S}_y \hat{S}_y \rangle}\right).
\end{equation}

\subsection{\label{sec:trapped_ion_connection}Connection to trapped ions}

Equations (\ref{eq: a_expectation}-\ref{eq: ab_correlations}) are very general, enabling a complete description of spin-spin correlations in a variety of different physical systems, but they cannot be further simplified or evaluated without choosing explicit forms for the boson spectrum $\omega_{\alpha}$ and spin-boson couplings $g_j^{\alpha}(t)$.  In the remainder of Sec.\ \ref{sec: spin_squeezing}, we compute the squeezing induced by the Hamiltonian in Eq.\,(\ref{eq: hamiltonian_for_ions}) using parameters relevant for ions in a linear Paul trap, though many of the qualitative features discussed below are insensitive to the detailed trap geometry, and would be similar for ions in a Penning trap if the axial modes were being driven \cite{britton12}. Specifically, we examine the dynamics of 20 ions, and assume that the wave-vector difference of the driving lasers ($\bm{k}_{\rm rel}$) points along one of the two transverse principal axes of the trap.  In this configuration, the spins only couple to normal modes oscillating along a single spatial direction, and therefore the number of (coupled) phonon modes is equal to the number of spins; $\mathcal{N} \equiv \mathcal{N}_{\rm s} = \mathcal{N}_{\rm b} = 20$. To calculate the normal-mode eigenvectors and associated frequencies, we set the ratio of longitudinal to transverse trap frequencies equal to $0.1$. The mass of the ions enters the calculation only as an overall scaling of the normal-mode frequencies, and can be ignored if all energies (times) are measured in units of the center-of-mass frequency $\omega_{\rm com}$ ($1/\omega_{\rm com}$).  The phonon modes are calculated in the standard fashion assuming a pseudopotential approximation holds \cite{James_1998,James_2003,PhysRevA.87.013422}, which neglects any effects due to micromotion. We first find the equilibrium positions of the $\mathcal{N}$ ions in the trap for the given trap curvatures in the different spatial directions and the mutual Coulomb repulsion of the ions. Then we determine the spring constant matrices by expanding the Coulomb interaction to quadratic order about equilibrium. Diagonalizing these dynamical matrices yields both the normal-mode oscillation frequencies $\omega_\alpha$ and the associated orthonormal eigenvectors $b^\alpha_j$.  Because the spring-constant matrices are real and symmetric, the $b_j^{\alpha}$ are real, and the spin-phonon coupling constants satisfy $g_{j}^{\alpha}(t)=\bar{g}_{j}^{\alpha}(t)=\Omega\sin(\mu t)\eta_{\alpha}b^{\alpha}_j$.

\begin{figure}
	\centering
	\includegraphics[width=0.98\columnwidth]{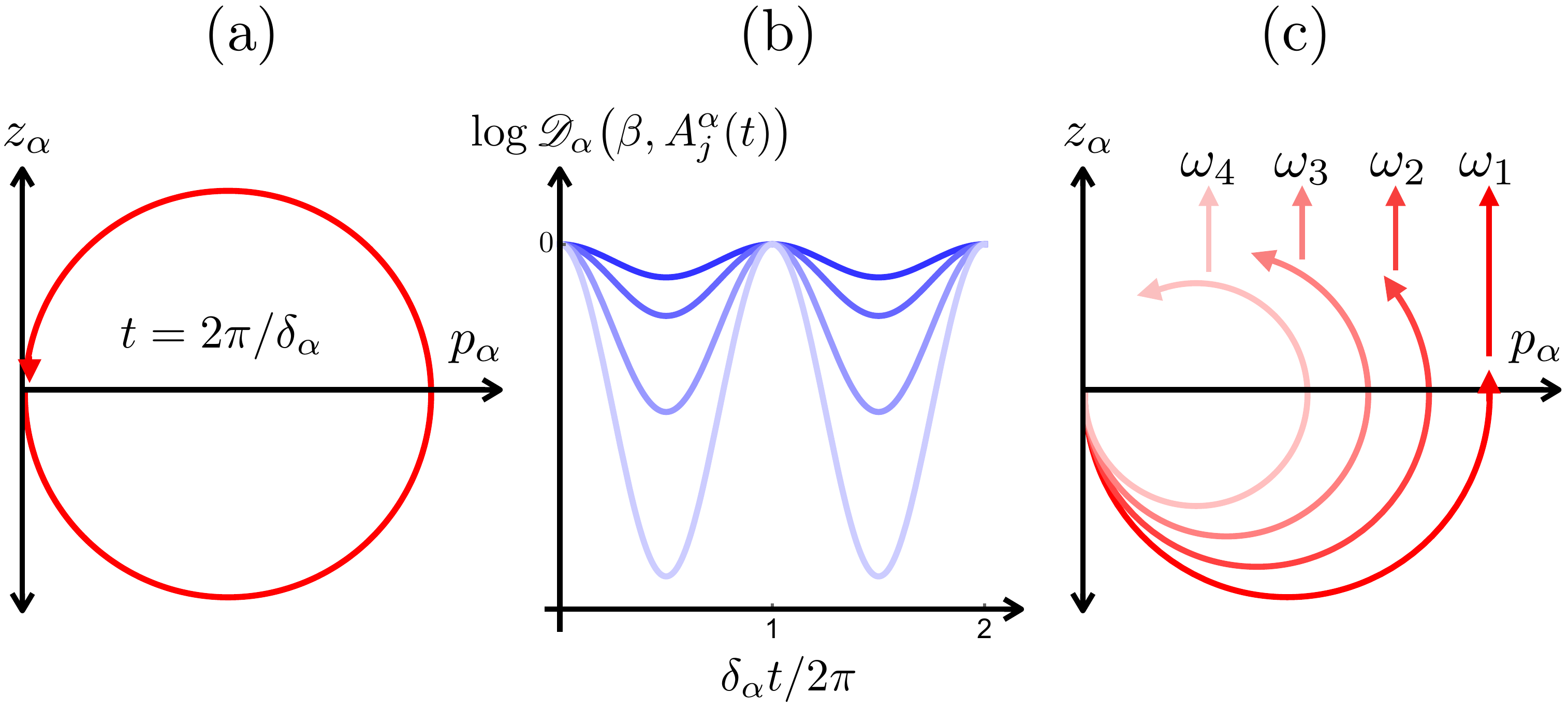}
	\caption{(Color online) Schematic phase-space dynamics of the phonon modes.  (a) When driven close to resonance, a single mode returns to its original position in phase space periodically (here $p_{\alpha}={\rm Im}[A_{j}^{\alpha}(t)]$ and $z_{\alpha}={\rm Re}[A_{j}^{\alpha}(t)]$). (b) The phonon modes periodically disentangle at finite temperature, as can be seen from the expectation values $\mathscr{D}_{\alpha}\big(\beta,A_j^{\alpha}(t)\big)$ returning to unity.  The overlap of a given mode with its initial value, however, decreases more sharply away from the recurrence times at higher temperature (here $k_B T=\hbar\omega_{\alpha}\times\{0,2,5,10\}$, from darker to lighter curves).  (c) When multiple modes are driven, those further from resonance make smaller but faster excursions through phase space, and in general the different modes do not simultaneously return to their initial states.}
	\label{phonons}
\end{figure}

Substituting this expression for $g_j^{\alpha}(t)$ into Eqs.\,(\ref{eq: A_def_general}) and (\ref{eq: S_def_general}) and performing the integrals, we obtain
\begin{align}
A_j^\alpha(t)&=
 \frac{i\Omega\eta_\alpha b_j^\alpha }{\omega_{\alpha}^2-\mu^2}\bigg(\mu-e^{i\omega_{\alpha} t}\left[\mu\cos(\mu t)-i\omega_{\alpha}\sin(\mu t)\right]\bigg),\nonumber\\
\label{eq: s_ions}
\mathcal{S}_{jk}(t) &= -\Omega^2\sum_{\alpha=1}^{\mathcal{N}}\frac{\eta_\alpha^2 b_j^\alpha b_{k}^\alpha}{\omega_\alpha^2-\mu^2}\biggr(\frac{\omega_\alpha t}{2}-\frac{\omega_\alpha}{4\mu}\sin(2\mu t)\\
&-\frac{\mu^2\cos(\mu t)\sin(\omega_\alpha t)-\mu\omega_\alpha\sin(\mu t)\cos(\omega_\alpha t)}{\omega_\alpha^2-\mu^2}\biggr).\nonumber
 \end{align}
Here $A_{j}^{\alpha}(t)$ is proportional to the interaction-picture phase space displacement of the $j$$^{\rm th}$ spin as a result of periodically driving the mode $\alpha$.  Because this driving is periodic, the displacement amplitudes $A_{j}^{\alpha}(t)$ have a simple structure.  In particular, for a single mode driven near resonance ($|\delta_{\alpha}|\ll\omega_{\alpha}$, with $\delta_{\alpha}\equiv\mu-\omega_{\alpha}$), it is straightforward to show that
\begin{align}
A_{j}^{\alpha}(t)\approx \frac{i}{2}\frac{\Omega\eta_{\alpha}b^{\alpha}_j}{\delta_{\alpha}}\big(e^{-i\delta_{\alpha}t}-1\big).
\end{align}
This amplitude traces a closed circle, vanishing at times such that $\delta_{\alpha} t=2n\pi$ (Fig.\,\ref{phonons}a), with $n$ an integer, at which the phonon matrix elements of the mode $\alpha$ in Eqs.\,(\ref{eq: a_expectation}-\ref{eq: ab_correlations}) all become equal to unity.  At these times, the mode $\alpha$ becomes unentangled with the spins, regardless of its initial state.  Physically, the independence on initial state reflects the independence of the period of a harmonic oscillator on its displacement, and implies that the periodic disentanglement of a given phonon mode occurs even at finite temperature, which was pointed out originally in Ref.\,\cite{PhysRevLett.82.1971}.  Formally, it is immediately apparent from Eq.\,(\ref{eq: finite_T_matrix_element}) that finite-temperature phonon expectation values of the form $\mathscr{D}_{\alpha}\big(\beta,A_{j}^{\alpha}(t)\big)$ will return to unity periodically for any inverse temperature $\beta$ (see Fig.\,\ref{phonons}b).  However, when multiple modes participate in the dynamics, they do not all disentangle at the same time, with further off-resonant modes exhibiting phase space excursions with smaller radii but larger frequencies, as in Fig.\,\ref{phonons}c.

In Sec.\,\ref{sec:results_T_is_0}, we produce plots of the squeezing parameter $\xi$ as a function of time using the exact solution and also using two useful approximations.  In the first approximation, we evolve our initial state with the spin-only time evolution operator
\begin{align}
\mathcal{U}_{\rm spin}(t)&=\exp\bigg( -i\sum_{j,k=1}^{\mathcal{N}} \mathcal{S}_{jk}(t)\hat \sigma_j^z\hat\sigma_{k}^z\bigg),
\end{align}
which amounts to ignoring spin-phonon entanglement.  This evolution is achieved by replacing $A_{j}^{\alpha}(t)\rightarrow 0$ in Eqs.\ (\ref{eq: a_expectation}-\ref{eq: ab_correlations}) while treating the $\mathcal{S}_{jm}(t)$ [which in reality depend implicitly on the $A_{j}^{\alpha}(t)$] as independent parameters, and then evaluating the $\mathcal{S}_{jk}(t)$ using Eq.\,(\ref{eq: s_ions}).  In the second approximation, spin-motion entanglement is ignored and the spin-spin couplings are replaced with their time averages, yielding time evolution under a time-independent Ising spin model
\begin{align}
\mathcal{U}^{\rm avg}_{\rm spin}(t)&=\exp\bigg( -i t \sum_{j,k=1}^{\mathcal{N}} \mathcal{S}^{\rm avg}_{jk}\hat \sigma_j^z\hat\sigma_{k}^z\bigg),
\end{align}
with time-averaged coupling constants
\begin{align}
\mathcal{S}_{jk}^{\text{avg}} =\lim_{t\rightarrow\infty}\frac{\mathcal{S}_{jk}(t)}{t}= -\frac{\Omega^2}{2}\sum_{\alpha=1}^{\mathcal{N}}\frac{\eta_{\alpha}^2 b_j^\alpha b_k^\alpha \omega_\alpha}{\omega_\alpha^2-\mu^2}.
\end{align}
We produce plots by varying $\Omega$, which controls the spin-phonon coupling strength, $\delta\equiv\mu-\omega_{\rm com}$, which is the difference between the two-photon detuning of the Raman lasers and the center-of-mass mode frequency, and also the phonon temperature $T$, which controls the initial number distribution in the phonon modes.  For the rest of the paper, we measure all energies in units of the transverse center-of-mass mode frequency $\omega_{\rm com}$ ($\hbar=1$), and measure all temperatures in units of $\omega_{\rm com}/k_B$, with $k_B$ Boltzmann's constant.

\subsection{\label{sec:results_T_is_0}Phonon effects at $T=0$}

%
%

\begin{figure}[!t]
    \centering
    \includegraphics[width=0.7 \columnwidth]{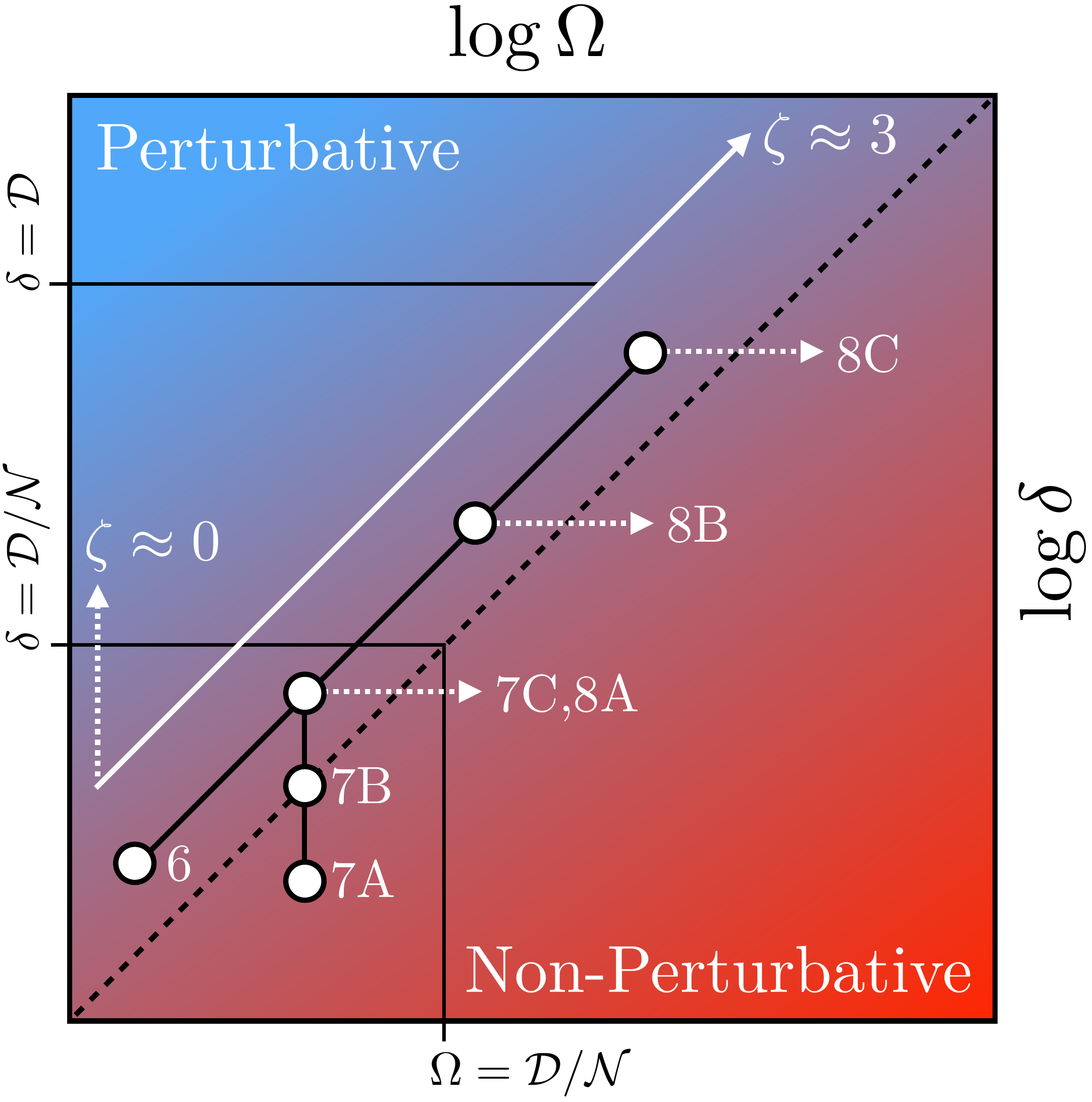}
    \caption{(Color online)
        The parameter space at zero temperature.  On this log scale, the ratio of $\Omega/\delta$, which controls the extent to which the system is in the perturbative limit, is given by the distance to the diagonal dashed line.  The boxed area on the lower left indicates the parameter space in which the center-of-mass mode dominates the dynamics.  Many trapped ion experiments attempt to operate in the perturbative limit, shown here as a white arrow.  Along this line, the decay of interactions can be tuned from $1/r^0$ (at $\delta\ll\mathcal{D}/\mathcal{N}$) to $1/r^{3}$ (at $\delta\gg\mathcal{D}$). The white dots with numeric labels indicate the parameters investigated in the numbered figures that follow.}
    \label{fig:parameter_space}
\end{figure}

Trapped ion experiments aimed at quantum simulations generally cool the phonon modes to temperatures on the order of, or ideally lower than, typical phonon energies.  However, in general, and especially for 2D or large 1D crystals, these temperatures are decidedly \emph{not} zero for all of the relevant modes.  Nevertheless, the phonons are often assumed to be cooled well enough that one can approximate them as being at zero temperature; we will address the zero-temperature situation first, and later come back to analyze the consequences of having thermal phonons in the initial state.  At zero temperature, the qualitative structure of the dynamics is determined by two independent considerations.  First, the coupling strength to a given phonon mode, measured relative to its detuning from the drive frequency $\mu$, determines how strongly that mode is driven.  Because the vibrations of the particles in the trap are transverse, all vibrational modes have frequency $\omega_\alpha \leq \omega_{\rm com}$. As a result, by choosing $\delta\geq0$ (so that no modes other than the center-of-mass mode can be resonant), we ensure that $\Omega/\delta$ is a suitable measure of how deeply in the ``perturbative'' limit the system is.  When this ratio is small, all phonon modes are weakly populated in the dynamics, and we expect spin-phonon entanglement to be unimportant.  Conversely, when this ratio is large, spin-phonon entanglement will be important, and we do not expect the approximations described above to agree well with the exact solution.   Next, we must consider the absolute size of $\delta$ and $\Omega$ relative to the mode bandwidth, denoted by $\mathcal{D}$, which controls the relative extent to which \emph{different} phonon modes participate in the dynamics.  For example, when $\delta$ and $\Omega$ are small compared to the typical mode spacing, $\delta ,\Omega\ll\mathcal{D}/\mathcal{N}$, spin dynamics occurs primarily due to coupling to the center-of-mass mode.  By increasing $\delta$ from much smaller than the typical mode spacing to much larger than the mode bandwidth, all the while keeping $\Omega\lesssim\delta$, one can navigate between two extreme limits: (A) For $ \delta\ll\mathcal{D}/\mathcal{N}$, the center-of-mass mode dominates the mediation of spin-spin interactions, which are therefore independent of the distance between two spins (since the center of mass mode is spatially homogeneous).  (B) For $\delta\gg \mathcal{D}$, all modes participate equally in mediating the spin-spin interactions, which fall off roughly as the cube of the distance between two spins \cite{PhysRevLett.92.207901}.  In between these extremes, it is common to approximate the spin-spin interaction to decay as a power law, $\mathcal{S}^{\rm avg}_{jk}\propto1/r^{\zeta}$, with $r$ the distance between ions $j$ and $k$ and $0<\zeta<3$.  All of these considerations are summarized in Fig.\,\ref{fig:parameter_space}, where a guide to the parameter space explored in the rest of this section can also be found.

\begin{figure}[t!]
    \centering
    \includegraphics[width=0.9 \columnwidth]{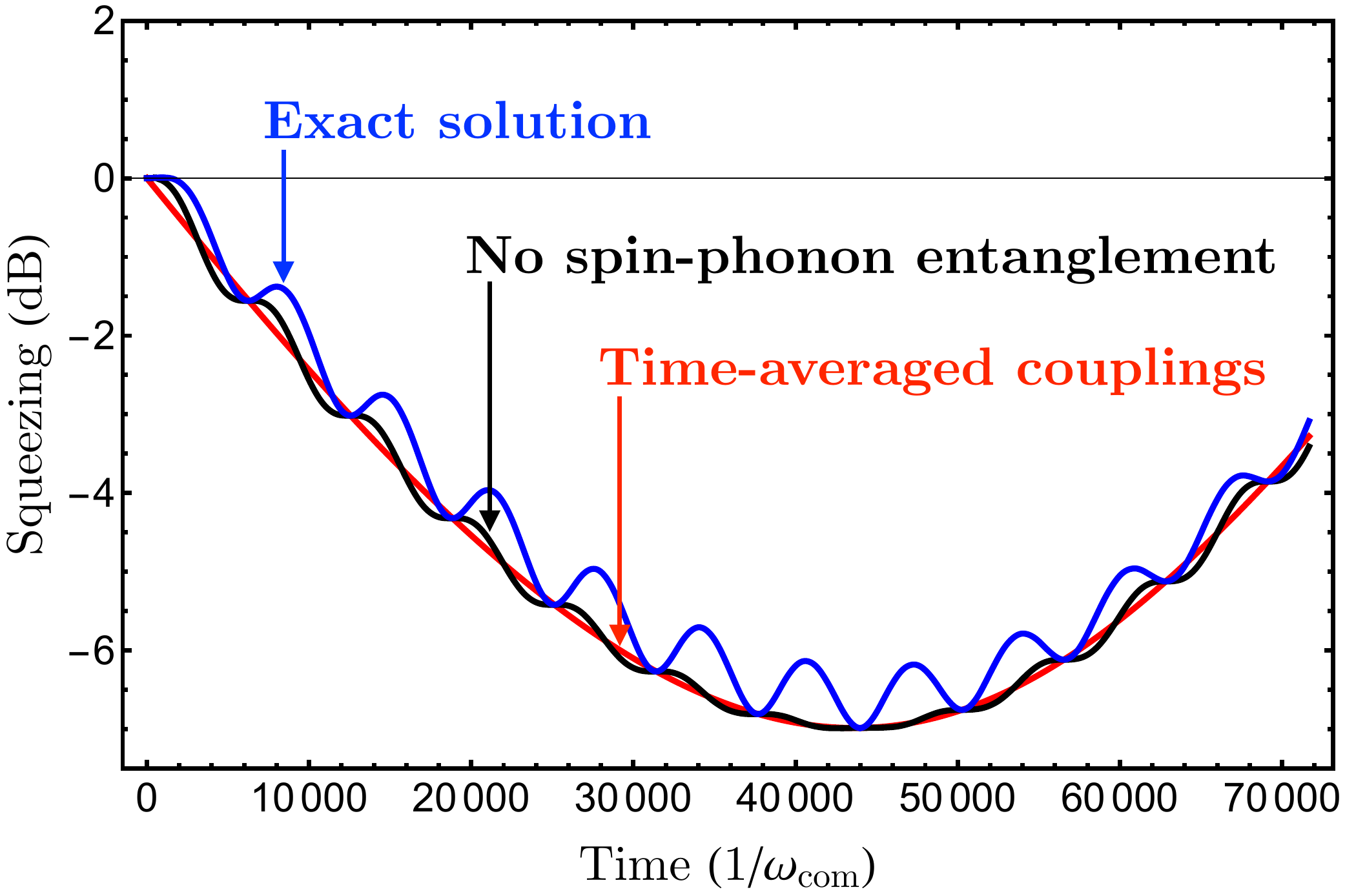}
    \caption{(Color online) Plots of the squeezing parameter $\xi$ as a function of time at zero temperature and with $\mu$ tuned very close to $\omega_{\rm com}$.  The parameters used here are $\{T=0$, $\delta =  10^{-3}$, $\Omega = 2.5\times 10^{-4}\}$.  The three curves show the results of the three different approximations described in Sec.\,\ref{sec:trapped_ion_connection}.  The blue curve is the exact solution, the black curve ignores spin-phonon entanglement but retains the full time-dependence of the spin couplings $\mathcal{S}_{jk}(t)$, and the red curve uses the time-averaged spin couplings $\mathcal{S}^{\rm avg}_{jk}$.}
    \label{initial_plot}
\end{figure}

\begin{figure*}[!t]
	\centering
	{\includegraphics[height=0.23\textwidth]{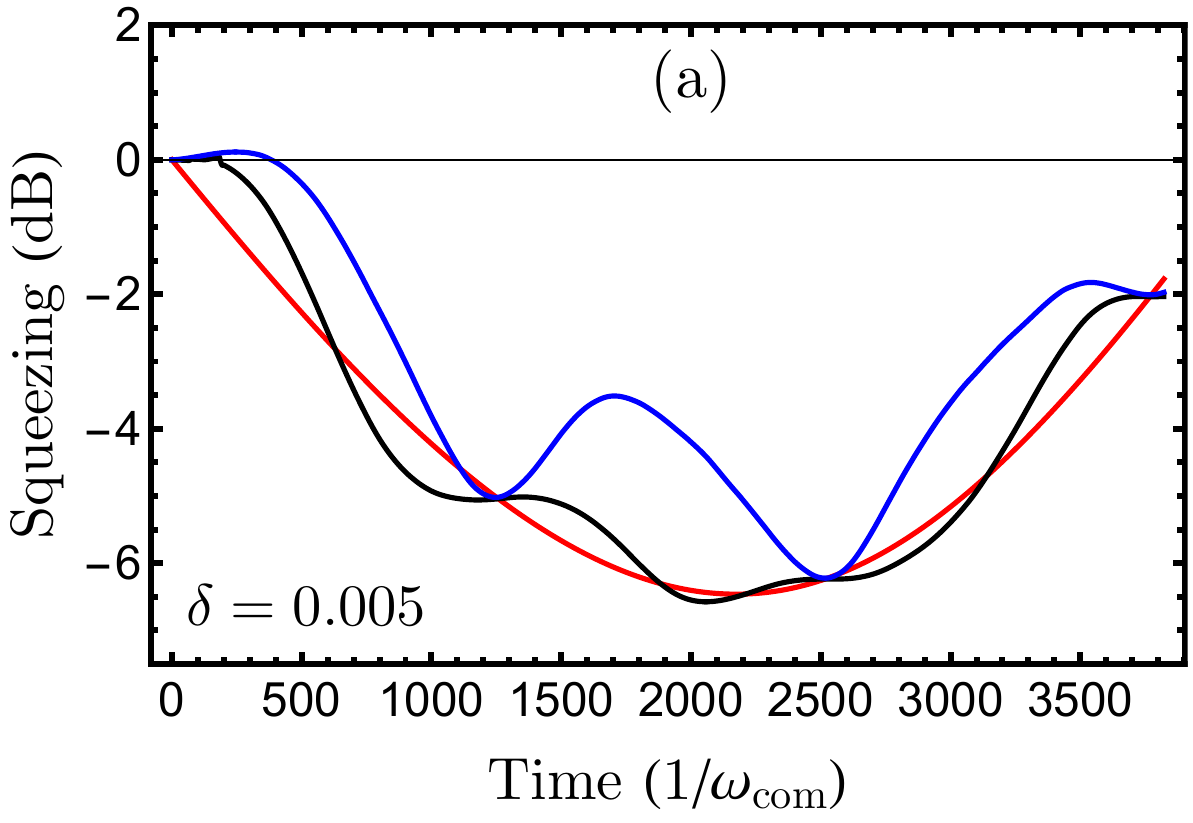}
		\label{varied_delta_const_omega_1}}
	{\includegraphics[height=0.23\textwidth]{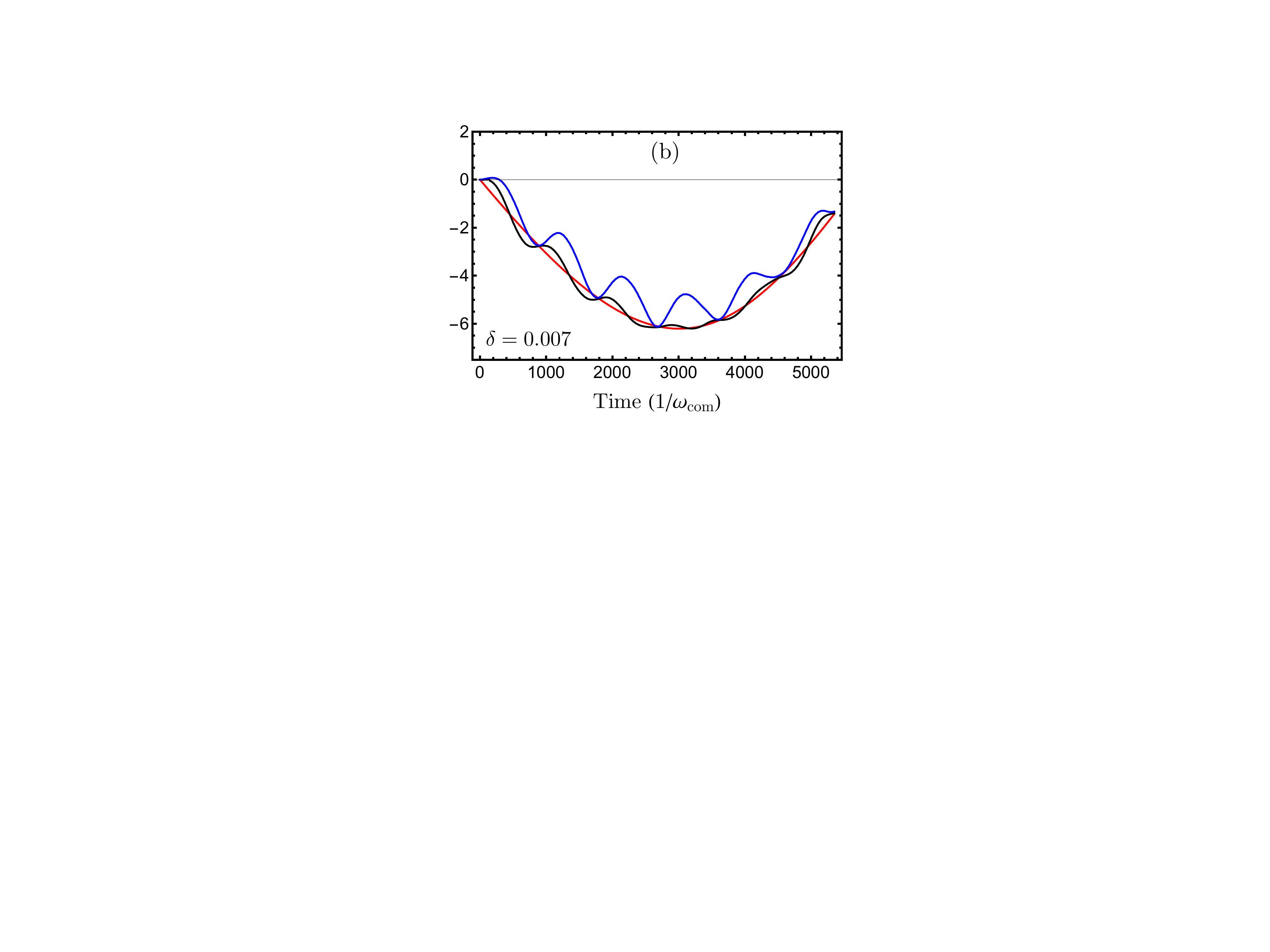}
		\label{varied_delta_const_omega_2}}
	{\includegraphics[height=0.23\textwidth]{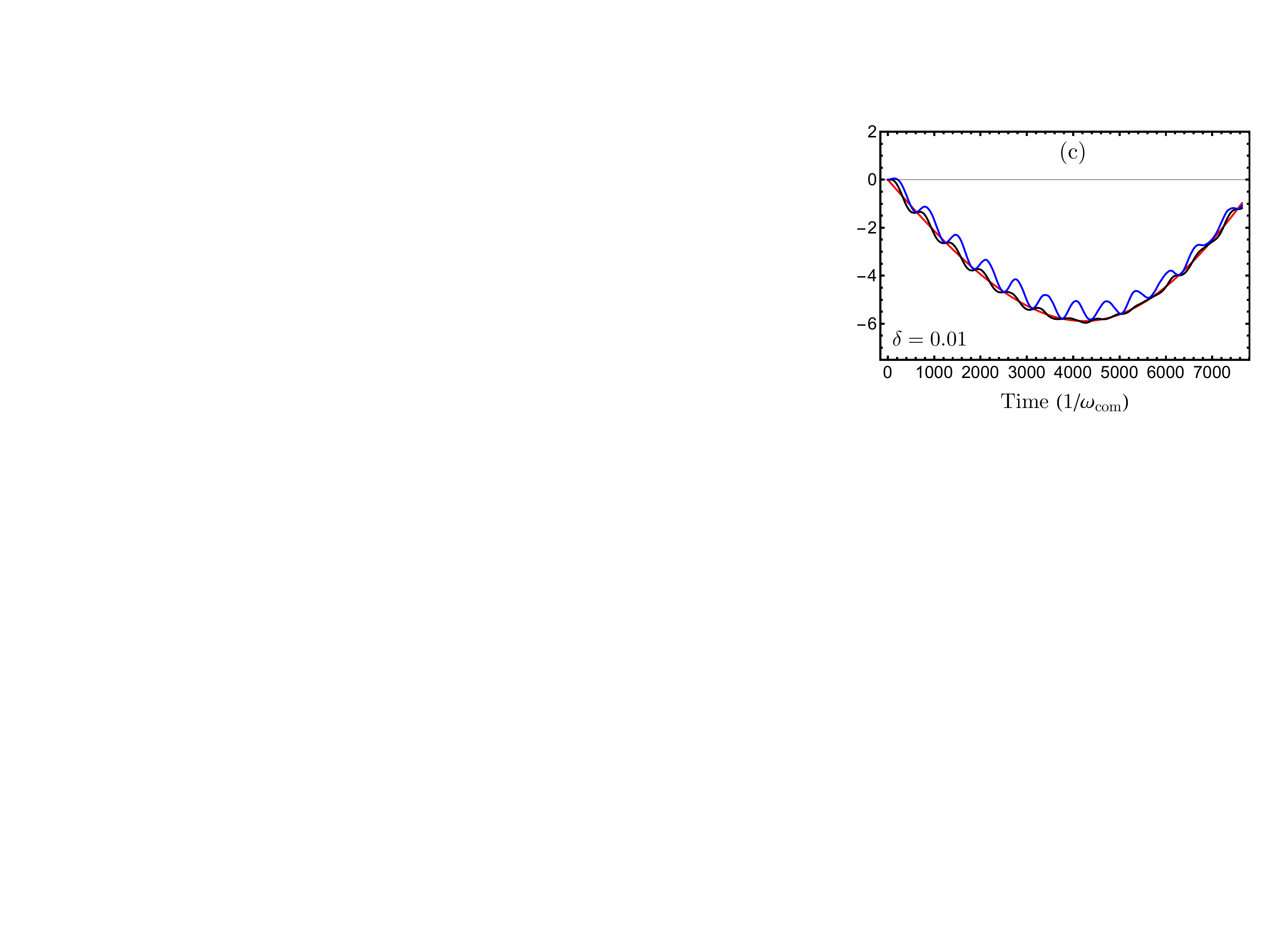}
		\label{varied_delta_const_omega_3}}
	\caption{(Color online) At zero temperature, the detuning from the center-of-mass mode is varied at constant $\Omega=2.5\times 10^{-3}$.  Note that the time-axis scaling is different in each plot, and the optimal squeezing occurs progressively later as $\delta$ is increased. The color code here is the same as that used in Fig.\,\ref{initial_plot}.}
	\label{varied_delta_const_omega_plots}
\end{figure*}

\begin{figure*}[!t]
	\centering
	{\includegraphics[height=0.225\textwidth]{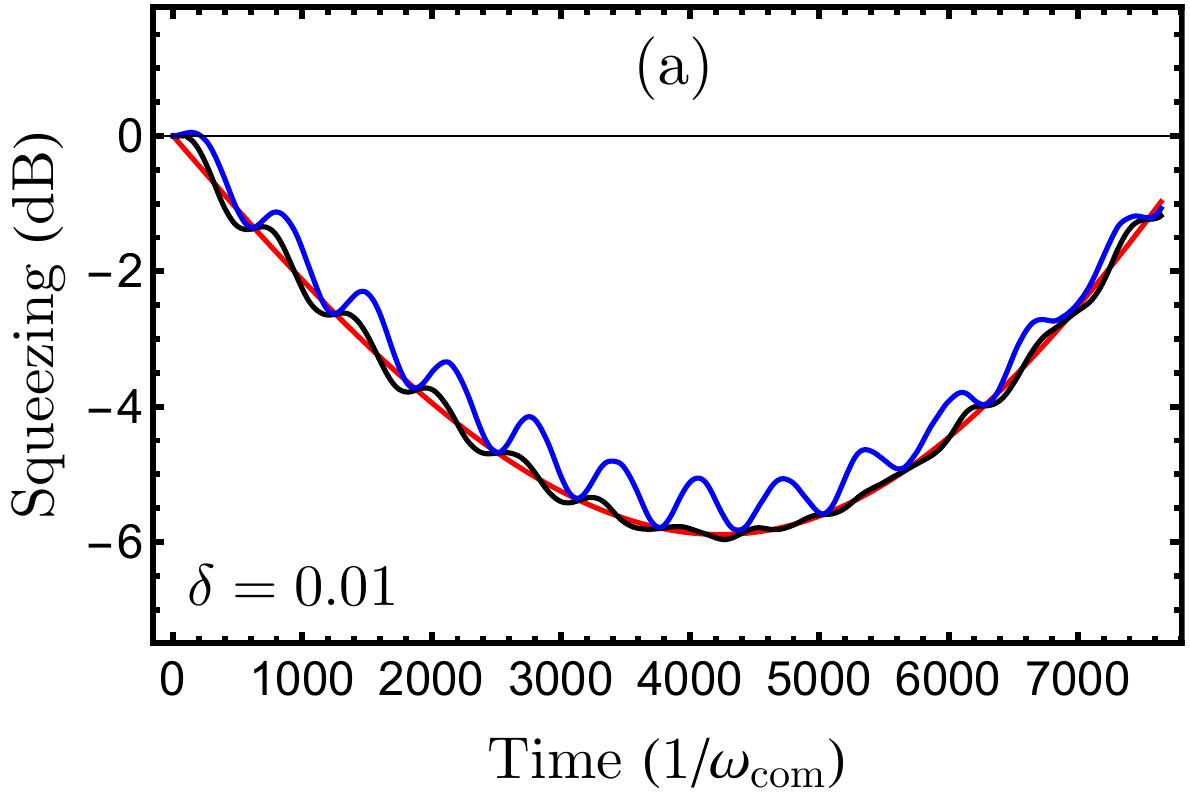}
		\label{varied_delta_1}}
	{\includegraphics[height=0.225\textwidth]{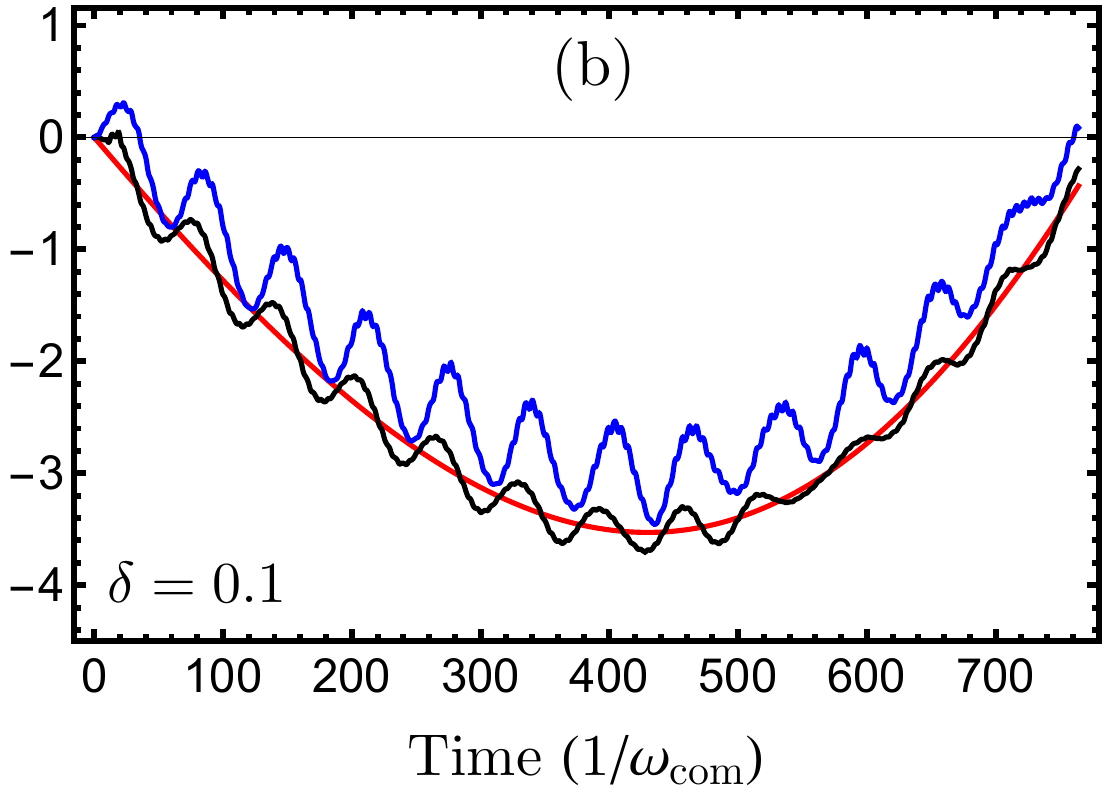}
		\label{varied_delta_2}}
	{\includegraphics[height=0.225\textwidth]{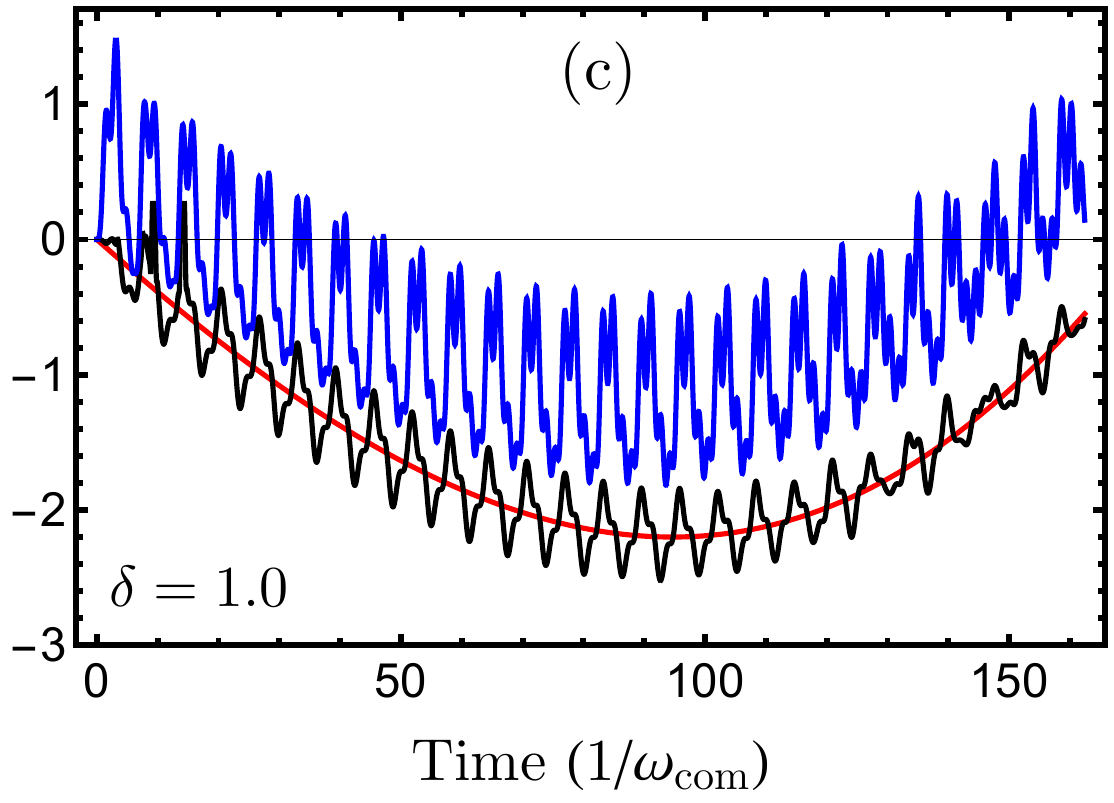}
		\label{varied_delta_3}}
	\caption{(Color online) At zero temperature, the detuning from the center-of-mass mode is varied while keeping $\Omega/\delta=1/4$, thereby controlling the relative contribution of the different phonon modes without greatly affecting the extent to which the individual modes are in the perturbative limit (i.e. $\Omega/(\mu-\omega_{\alpha})$ is not changing very much).  Note that as $\delta$ grows larger, the time of optimal squeezing becomes shorter and the total amount of squeezing obtained at that time is shrinking. The color code here is the same as that used in Fig.\,\ref{initial_plot}.}
	\label{varied_delta_plots}
\end{figure*}

Figure \ref{initial_plot} illustrates the behavior of the squeezing parameter $\xi$ at $T=0$, with the detuning chosen close to the center-of-mass mode ($\delta\ll\mathcal{D}/\mathcal{N}$) and marginally in the perturbative limit ($\Omega/\delta=1/4$). The squeezing parameter is normalized such that $\log(\xi)=0$ for a coherent state, so the region of the plot where the squeezing parameter dips below the horizontal axis denotes a period of improved uncertainty with respect to the standard quantum limit.  Under these conditions, the exact solution and the two approximations exhibit fairly similar behavior.  The smooth red curve produced by the time-averaged spin-coupling approximation correctly captures the overall trend of squeezing observed in the exact solution.  As seen in the black curve, the time-dependence of the spin couplings produces small amplitude 
high-frequency
oscillations. These oscillations are amplified by spin-phonon entanglement, as indicated by the exact solution (blue curve), showing that even at zero temperature and nominally in the perturbative limit, the creation of phonons during the dynamics can significantly affect the spin squeezing.   We note that very slight improvements in squeezing over the time-averaged spin-coupling approximation do periodically occur. They appear to be due to the time dependence of the true spin couplings $\mathcal{S}_{jk}(t)$, rather than the spin-phonon entanglement, as they occur in both the blue and black curves (and the latter bounds the former from below).
\newline
\indent To better understand the role that dynamical phonon creation plays in spin squeezing, we change the detuning from the center-of-mass mode, $\delta$, while holding $\Omega$ constant. Figure \ref{varied_delta_const_omega_plots} illustrates a series of results in which $\delta$ is increased but kept small compared to the detuning from all other modes (\emph{i.e.}\ $\delta\ll\mathcal{D}/\mathcal{N}$). These plots therefore primarily reflect changes in behavior caused by variations in the strength with which the COM mode is driven relative to its detuning, \emph{i.e.}\ the ratio $\Omega/\delta$.  In the time-averaged spin-coupling approximation (red curve), the dynamics is completely insensitive to this ratio up to an overall timescale, and upon scaling the maximum time by $\delta/\Omega$, we observe nearly the same squeezing behavior in all three panels.  In the first plot, $\delta/\Omega$ is relatively small; the center-of-mass mode is strongly driven, resulting in strongly oscillatory spin couplings and large spin-phonon entanglement.  In this limit, neither approximation accurately captures the exact dynamics, nor do they agree with each other.  As $\delta$ is increased, the oscillation amplitudes of the time-dependent couplings $\mathcal{S}_{jk}(t)$ diminish compared to their time-averaged values, so the two approximations begin to agree better with each other. At the same time, population in the center-of-mass mode becomes suppressed, spin-motion entanglement diminishes, and the exact result begins to converge to both approximations.  As discussed in Sec.\,\ref{sec:trapped_ion_connection}, when considering coupling only to the center-of-mass mode, the phonon degree of freedom periodically becomes unentangled from the spins, even at strong driving.  This behavior is reflected in the periodic agreement between the exact solution and the approximation ignoring spin-phonon entanglement.
\newline
\indent In all panels of Fig.\,\ref{varied_delta_const_omega_plots}, both $\delta$ and $\Omega$ are kept small compared to the mode spacing, and therefore the variation of $\Omega/\delta$ is the dominant factor contributing to the changes in behavior.  However, these plots provide little insight into the other important effect of increasing the detuning $\delta$: the increased importance of modes other than the center-of-mass mode.  In order to isolate the latter effect, in Fig.\,\ref{varied_delta_plots} we again vary $\delta$ but now keep the ratio $\Omega/\delta=1/4$ fixed.  This keeps the coupling to any given mode in the (barely) perturbative limit, thus counteracting the dominant role that varying spin-phonon entanglement played in the qualitative trends observed in Fig.\,\ref{varied_delta_const_omega_plots}.  As $\delta$ is increased, the time-averaged spin-coupling approximation correctly captures an overall trend of the exact solution: The time of optimal spin squeezing is becoming shorter, and the extent of squeezing that occurs at that time is diminishing.  The former effect is simply the result of increasing the spin-phonon coupling, which increases the overall rate of dynamics.  The reduction of squeezing achieved at the optimal time reflects the diminishing spatial-range of the spin-spin couplings (approaching $1/r^3$ for large $\delta$) as more phonon modes participate in mediating them.  With or without the inclusion of spin-phonon entanglement (i.e.\ in the blue or black curves), the squeezing exhibits high-frequency oscillations arising from the coupling to phonon modes other than the center-of-mass mode.  As multiple phonon modes at different frequencies become entangled with the spins, it is no longer possible for \emph{all} of those modes to become disentangled from the spins simultaneously.  This is strongly reflected in the third panel, where the exact solution no longer agrees with either approximation at regular intervals.

\subsection{Effects of finite temperature}

In Sec.\,\ref{sec:results_T_is_0}, we were primarily interested in the effects of dynamical phonon generation starting from the phonon vacuum.  In many experiments, especially those using large numbers of ions, this starting state is not realistic.  For example, in Ref.\,\cite{britton12}, the Doppler-cooling limit of $T\sim 1{\rm mK}\gtrsim 10\times( \hbar \omega_{\rm com}/k_{B})$ leads to $\gtrsim 10$ phonons per transverse mode.  As explained in Sec.\,\ref{sec:phonon_matrix_elements}, spin-spin correlation functions can also be computed at finite temperatures, and the above analysis can therefore be extended to capture the consequences of non-zero initial motional temperature on spin squeezing.

\begin{figure}[!t]
	\centering
	\includegraphics[width=0.75 \columnwidth]{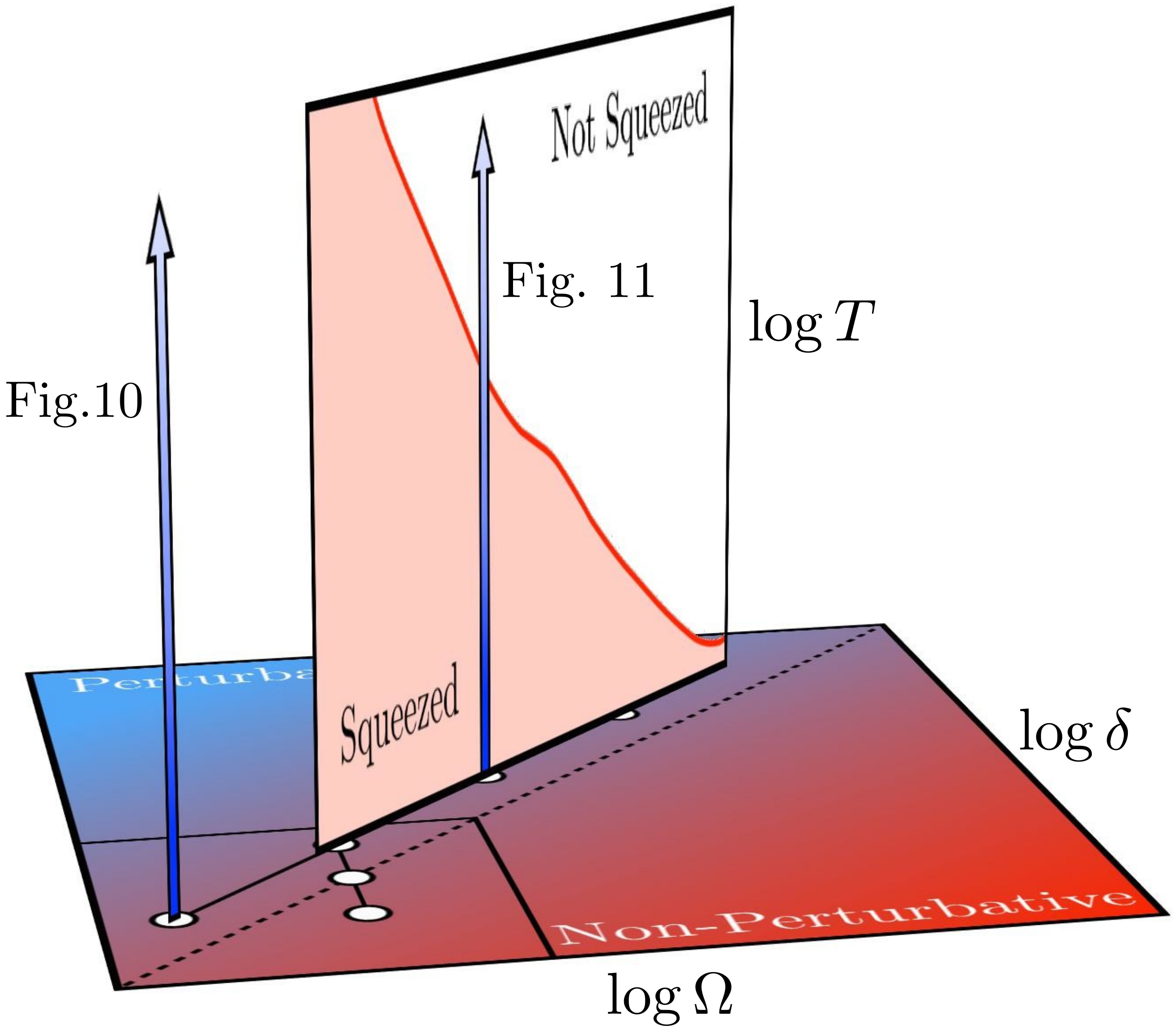}
	\caption{(Color online) Schematic illustration of a two-dimensional cross-section of the full three-dimensional parameter space spanned by $\Omega,\delta$, and $T$.  Working just barely in the perturbative limit (and at a fixed value of $\Omega/\delta$), we explore the effects of temperature for multiple values of $\delta$, which controls the relative participation of the various phonon modes in mediating spin-spin interactions.  The parameter space explored in Fig.\,\ref{varied_T_plots} and Fig.\,\ref{varied_delta0_and_T_plots} are indicated as vertical arrows.}
	\label{fig:T_delta}
\end{figure}

\begin{figure}[!t]
	\centering
	{\includegraphics[width=0.9\columnwidth]{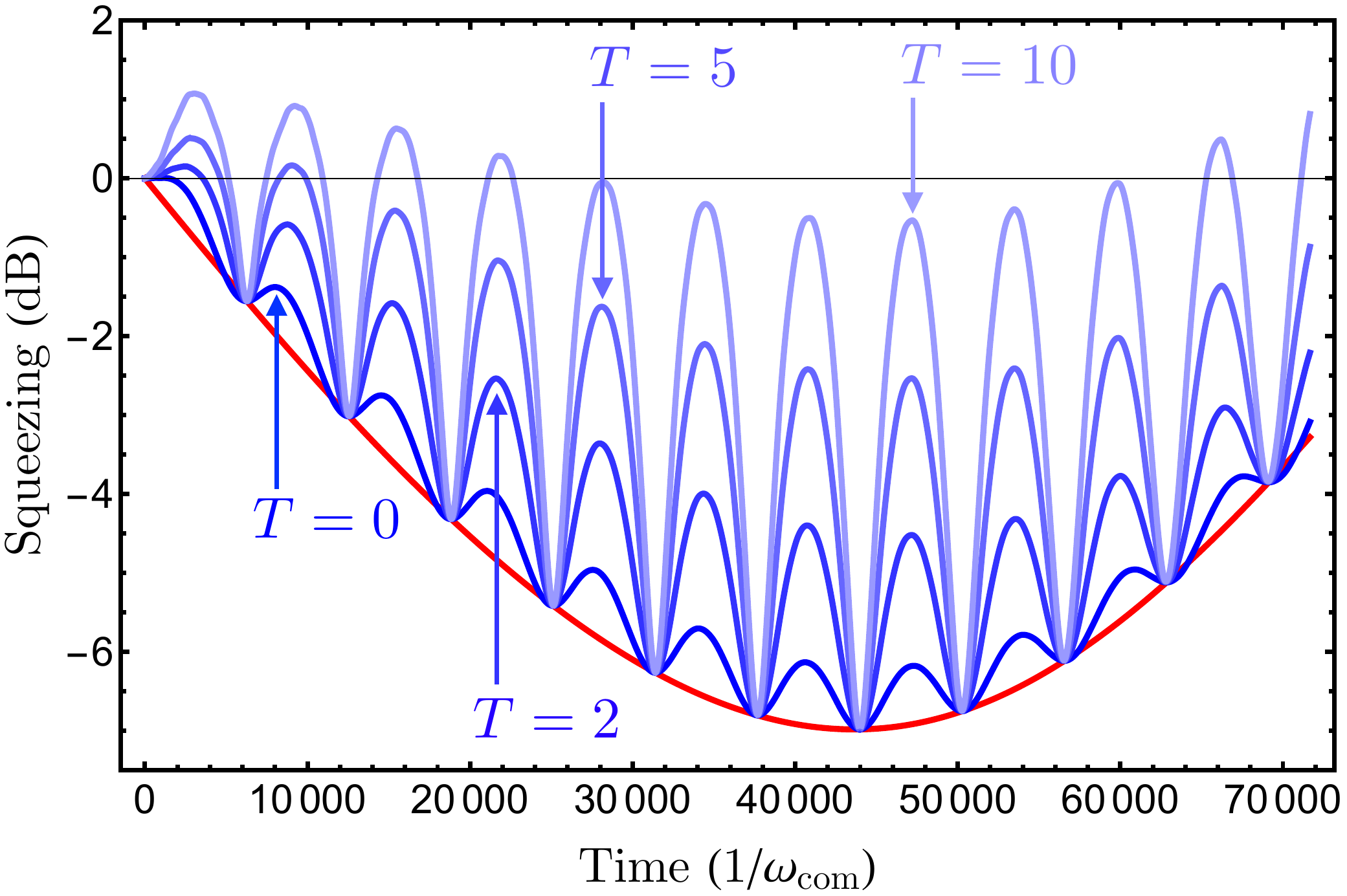}
		\label{varied_T}}
	\caption{(Color online) For a system driven very close to the center-of-mass frequency ($\delta = 0.001$, $\Omega = 2.5\times 10^{-4}$), spin squeezing is very robust against large initial temperatures.  Here the temperature is varied from zero up to $10\omega_{\rm com}$ ($\approx 0.1 \text{mK}$), and to a very good approximation the squeezing obtains its $T=0$ value at regular intervals.}
	\label{varied_T_plots}
\end{figure}

With the addition of temperature, there is a large parameter space to be explored; here we focus our attention on the barely perturbative limit at fixed $\Omega/\delta=1/4$ (the same ratio used in Fig.\,\ref{varied_delta_plots}), and consider both variations of $\delta$ and $T$ (see Fig.\,\ref{fig:T_delta} for a guide to the parameter space explored).  We first examine the case of near-detuning from the center-of-mass mode ($\delta,\Omega\ll\mathcal{D}/\mathcal{N}$), but now taking the phonon modes to be at a temperature $T$ at time $t=0$ (Fig.\,\ref{varied_T_plots}).  Here we plot just the exact solution and the time-averaged spin-coupling approximation (both the time-averaged spin-coupling approximation and the approximation of ignoring spin-phonon entanglement are insensitive to the phonon distribution, so neither varies with $T$). The primary effect of increasing temperature on the squeezing is that the amplitude of oscillations above the curve obtained from the time-averaged spin-coupling approximation increases. Nevertheless, the exact solution continues to agree with the approximation at regular intervals.  As discussed in Sec.\,\ref{sec:trapped_ion_connection}, this behavior can be understood as the insensitivity of a harmonic oscillator's period to its state of excitation; at finite temperature many Fock states of the center-of-mass mode are occupied, but as they are driven periodically they all return to their initial point in phase space simultaneously, at which point they are unentangled from the spins.  As $T$ becomes larger, the system spends most of its time in states with large uncertainty (poor squeezing), but precisely timed measurements of the system could nevertheless yield a significantly improved resolution over the standard quantum limit.  Interestingly, at times short compared to $2\pi/\delta$, the spin distribution is \emph{always} antisqueezed, and the degree of antisqueezing could in principle be used to perform in-situ temperature measurements of the phonon modes.

\begin{figure}[!t]
	\centering
	{\includegraphics[width=0.9 \columnwidth]{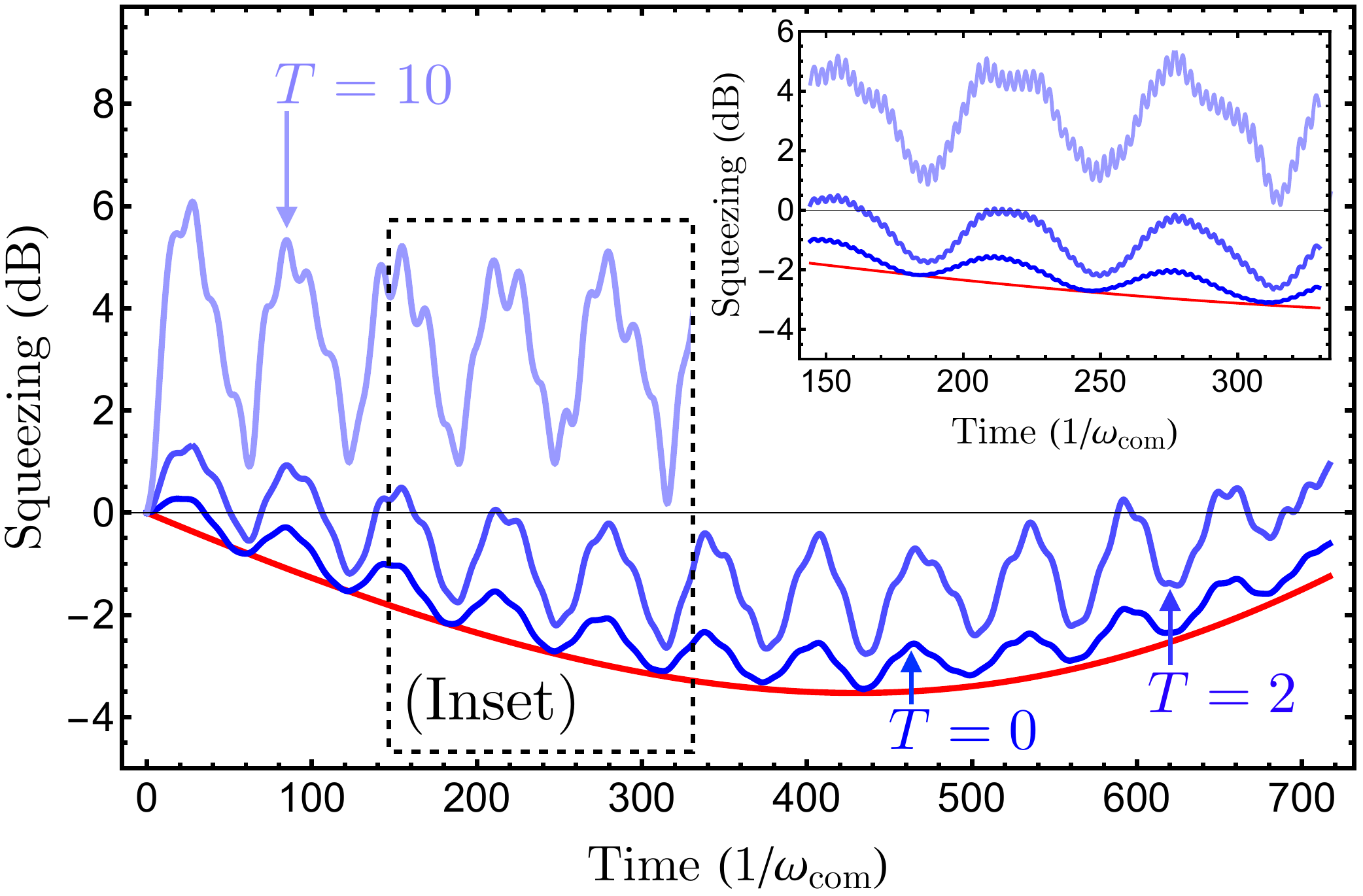}
		\label{varied_T_detuned}}
	\caption{(Color online) For a system driven further off resonance from the center-of-mass frequency ($\delta=0.1$, $\Omega = 0.025$), finite temperature has a much more detrimental effect on the spin squeezing, eventually preventing any squeezing from occurring beyond a critical temperature (here $T\approx 10$). Inset: Note that, due to the relatively large population of many modes, there is significant high frequency structure that is hidden on the time scale over which squeezing occurs.}
	\label{varied_delta0_and_T_plots}
\end{figure}

In Fig.\,\ref{varied_delta0_and_T_plots}, we explore the behavior of the system detuned away from the center-of-mass frequency and at finite temperature.  The behavior exhibited in Fig.\,\ref{varied_delta0_and_T_plots} reflects the general trends observed in the previous plots: increasing $\delta$ induces additional high-frequency structure in the exact solution, and increasing $T$ produces an overall growth in the amplitude of oscillations in the spin squeezing. Unlike in Fig.\,\ref{varied_T_plots}, however, the exact solution not only shows increased oscillation amplitude at higher temperature, but in addition the local squeezing minima are also displaced increasingly far from the curve calculated in the time-averaged spin-coupling approximation. Even though at $T=0$ the squeezing nearly agrees with this approximation throughout the entire time region plotted, spin squeezing is completely destroyed when the temperature reaches $T\approx 10$.

\begin{figure}[!t]
\centering
\includegraphics[width=0.9\columnwidth]{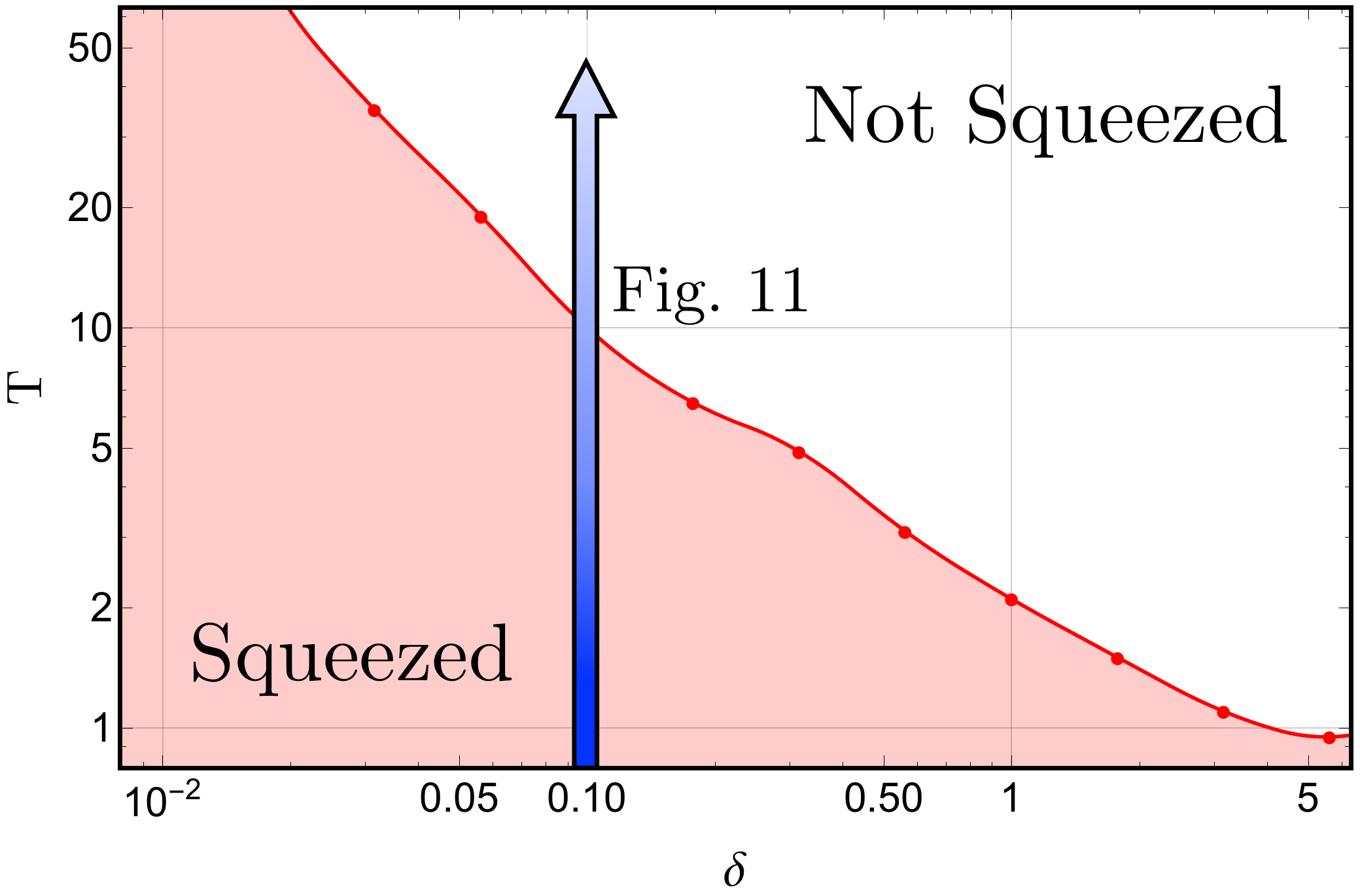}
\caption{(Color online) Squeezing ``phase-diagram'' in the $\delta$-$T$ plane, showing that no squeezing occurs above a critical temperature (note that $\Omega/\delta$ is being held constant as $\delta$ is changed, see Fig.\,\ref{fig:T_delta}).}
\label{spin_squeezing_phase_boundary}
\end{figure}

Indeed, for a given detuning there will always be some temperature threshold above which no squeezing takes place at all ($\xi\geq 1$ for all $t$). By varying $T$ at evenly spaced intervals of $\log\delta$, in Fig.\,\ref{spin_squeezing_phase_boundary} we produce a ``phase diagram'' demarcating this boundary in the parameter space.  The qualitative downward trend of the boundary can be understood as an increased sensitivity of the spin dynamics to the initial phonon temperature as more phonon modes participate in the dynamics. The additional modes prevent the spins from becoming periodically unentangled with the phonons; the consequences of this residual spin-motion entanglement on spin squeezing are exacerbated at finite temperature because the occupation of a given phonon mode affects the extent to which that mode remains entangled with the spins at a given time (see Fig.\,\ref{phonons}b).  At small $\delta$, squeezing persists even for temperatures on the order of 100 times the center-of-mass-mode energy.  However, note that at these high temperatures the approximation that the phonons are non-interacting (equivalently that the potential seen by an individual ion is harmonic) is likely to break down, and the system may well be outside of the Lamb-Dicke limit used to justify a description in terms of Eq.\,(\ref{eq: Hamiltonian}).

\section{\label{sec:outlook}Discussions and Future Directions}

While Eq.\,(1) is quite general, and can describe a diverse array of practically relevant experiments, there is certainly a sense in which it is extremely constrained---it has a large number of locally conserved quantities.  Indeed, a careful inspection of our calculations reveals that the local conservation of $\hat{\sigma}^z_j$ is, at several different levels, directly responsible for the solvability of the model.  Nevertheless, our solutions form a useful benchmark for powerful (but computationally expensive) tools capable of numerically solving the non-equilibrium dynamics of more general models, such as time-dependent versions of matrix-product-state algorithms \cite{Schollwock_2011}.  Especially in 2D---but generally even in 1D given the long-range nature of the spin-couplings induced by the (often) delocalized bosonic modes---it remains unclear to what extent there are \emph{any} general purpose and controlled numerical tools for studying this dynamics.  We also point out that questions about equilibration and thermalization, which require studies of \emph{long-time} dynamics, are difficult even in 1D for all but the smallest systems.

The exact solution will be useful in testing a variety of experimental idealizations that are frequently made but generally not quantitatively justified.  For example, many trapped-ion experiments employ a spin-echo pulse in order to obtain a coherence time on the order of the spin-spin interaction time.  If the spins are unentangled with the phonons, then a spin echo completely removes the effects of an inhomogeneous magnetic field ($\sim \sum_{j}h_j\hat{\sigma}_j^z$) added to the Hamiltonian in Eq.\,(\ref{eq: Hamiltonian}).  However, the existence of spin-phonon entanglement at the time of the echo pulse invalidates this picture \cite{PhysRevA.92.033608}, and the consequences on spin-spin correlation functions could be explored using the solution developed in this paper.

There are also many interesting purely theoretical questions about the present model, many of which we believe the tools developed here are well suited to begin answering.  For example, while spin-squeezing calculations reveal the impact of spin-boson entanglement on attempts to generate strictly spin-spin entanglement, it should also be possible to analyze spin-boson entanglement more directly by calculating correlation functions involving both spin and boson operators.  It would also be interesting to explore to what extent the ``solvability'' of this model can be extended to calculating more general quantities than low-order correlation functions.  While our solution enables efficient calculation of arbitrary-order correlation functions of the form given in Eqs.\,(\ref{eq: cf_1}-\ref{eq: cf_3}), calculating the full counting statistics along any spin direction orthogonal to $z$ is still very difficult.  In particular, computation of the full counting-statistics is equivalent to the computation of $\sim\mathcal{N}^{\rm th}$-order correlation functions of Pauli ($x,y,z$) matrices, which involves the summation of an exponentially large (in $\mathcal{N}$) number of high-order correlation functions of the form in Eqs.\,(\ref{eq: cf_1}-\ref{eq: cf_3}).  It seems very plausible that this difficulty is related to computational hardness results for classically sampling spin distributions following dynamics under commuting spin Hamiltonians \cite{Bremner_2010}.  The present model adds an interesting twist, in that it builds bosonic degrees of freedom into a commuting spin Hamiltonian in a way that preserves its solvability (in the sense of calculating low-order correlation functions); the bosons alone, despite being non-interacting and therefore ``solvable'' in the same sense as the model studied here, are thought to be hard to simulate clasically in a precise sense \cite{Aaronson_2011}.
%
%
%
%
%
%
%
%
%
%
%
%
%
\acknowledgments
We acknowledge helpful conversations with John Bollinger, Joe Britton, Zhexuan Gong, Bryce Yoshimura, Alexey Gorshkov, Bill Fefferman, and Arghavan Safavi-Naini.  J.K.F. acknowledges support from the National Science Foundation under grant number PHY-1314295 and from the McDevitt bequest at Georgetown University.  A.M.R. acknowledges funding from the NSF (PHY-1521080 and PFC-1125844), AFOSR, AFOSR-MURI, NIST, and ARO individual investigator awards.   M.L.W. and M.F.-F. thank the NRC postdoctoral fellowship program for support.



\end{document}